\documentclass[aps,pre,twocolumn,showpacs]{revtex4-1} 

\usepackage{lineno,hyperref}
\usepackage[T1]{fontenc}
\usepackage{lmodern}
\usepackage{microtype}
\usepackage{color}
\usepackage{tabularx}
\usepackage{booktabs}
\usepackage{multirow}
\usepackage[
	format=hang,
	indention=-1.0cm,
	justification=RaggedRight,
	font=small
]{caption}
\usepackage{graphicx} 
\usepackage{dcolumn} 
\usepackage{bm} 
\usepackage[version=3]{mhchem}
\usepackage{stackrel}

\usepackage{verbatim}   
\usepackage{subfigure}  
\usepackage[english]{babel}
\usepackage{calc}

\usepackage{amsmath, amsfonts, amssymb,latexsym}

\newlength\myheight
\newlength\mydepth
\settototalheight\myheight{Xygp}
\newcommand{\be}{\begin{equation}}
\newcommand{\ee}{\end{equation}}
\newcommand{\bea}{\begin{eqnarray}}
\newcommand{\eea}{\end{eqnarray}}

\newcommand{\BEQ}{\begin{equation}}     
\newcommand{\BEA}{\begin{eqnarray}}
\newcommand{\BD}{\begin{displaymath}}
\newcommand{\EEQ}{\end{equation}}       
\newcommand{\EEA}{\end{eqnarray}}
\newcommand{\ED}{\end{displaymath}}

\newcommand{\del}{\delta}

\newcommand{\eps}{\varepsilon}          

\newcommand{\half}{{1\over 2}}     

\newcommand{\R}{\mathbb{R}}

\newcommand{\Medskip}{\medskip\noindent}
\newcommand{\Bigskip}{\bigskip\noindent}
\newtheorem{Theorem}{Theorem} 
\newtheorem{Lemma}{Lemma} 

\begin{document}

\title{Phase transitions in optimal betting strategies}

\author{L. Dinis$^1$, J. Unterberger$^2$ and D. Lacoste$^3$}

\affiliation{                    
  $^1$ GISC - Grupo Interdisciplinar de Sistemas Complejos and Dpto. de Estructura de la Materia, \\
  Física Térmica y Electrónica, Universidad Complutense de Madrid, 28040 Spain \\
  $^2$ Institut Elie Cartan, UMR CNRS 7502, Universit\'e de Lorraine, \\ 
  BP 239 F-54506
Vandoeuvre-l\`es-Nancy Cedex, France \\
  $^3$ Gulliver Laboratory, UMR CNRS 7083, PSL Research University,  \\
ESPCI, 10 rue Vauquelin, F-75231 Paris Cedex 05, France
}



\date{\today} 

\begin{abstract}
Kelly's criterion is a betting strategy that maximizes the long term growth rate, but which is known to 
be risky. Here, we find optimal betting strategies that gives the highest capital growth rate while keeping a certain low value of risky fluctuations.  
We then analyze the trade-off between the average and the fluctuations of the growth rate, in models of horse races, 
first for two horses then for an arbitrary number of horses, and for uncorrelated or correlated races. 
We find an analog of a phase transition with a coexistence between two optimal strategies, 
where one has risk and the other one does not. 
The above trade-off is also embodied in a general bound on the average growth rate, similar to 
thermodynamic uncertainty relations. We also prove mathematically the absence of other 
phase transitions between Kelly's point and the risk free strategy.
\end{abstract}

\maketitle

\section{Introduction}

Developed in 1956 by Bell Labs scientist John Kelly, Kelly's criterion
applied the newly created field of information theory to gambling and
investment \cite{Kelly1956}. 
Largely popularized in books \cite{Poundstone2005}, 
this criterion allows a gambler (or investment fund) to fix 
what proportion of bankroll should be risked on a given bet. It
essentially exploits side information to maximize 
the expected geometric growth rate of a capital.
This work was precursor to the growth
optimal portfolio theory, which applied these ideas to 
capital market \cite{Thorp2011}. 
The ensemble of optimal investment strategies 
forms an efficient border \cite{Markowitz1952},  
or equivalently a Pareto front \cite{Seoane2015,Shoval2012}, which is a term used in engineering and economics to call the set of designs that represent best trade-offs between different conflicting requirements.

Recently, there has been a surge of interest in applying 
insights from optimal gambling theory and economy to
biology.  Kelly's work led to an essential clarification of the concept of   
 information value in biology
\cite{Bergstrom2004,Rivoire2011}, which was very helpful to understand strategies
used by biological systems in a fluctuating environment. In particular the
bet-hedging strategy turned out to be precisely an optimal strategy of the Kelly type \cite{Kussell2005,tal_adaptive_2020}.

Here, we focus on betting strategies of Kelly's type and draw inspiration from the field of Stochastic Thermodynamics, a recent branch of Thermodynamics with deep links
to information theory, and with already several works specifically applied to gambling or betting problems \cite{Neri2019,Ito2016,Vinkler2016,hirono_jarzynski-type_2015}.
A recent and an active line of research concerns the 
thermodynamic uncertainty relations
\cite{Horowitz2019,Falasco2019,Vroylandt2018,Proesmans2018}, which capture important tradeoffs in Thermodynamics. In this letter, we explore novel implications of these ideas for gambling models. 
We emphasize at this point that a background on Stochastic Thermodynamics is not required to understand this letter, 
since we only rely on basic notions of probability and optimization theory.

To gain insight into the tradeoff present in gambling,
we study the efficient border of Kelly's model, and we find that it extends to a region of negative growth, never discussed in the literature to our knowledge, corresponding to catastrophic betting strategies. 
Inspired by works on optimal protocols \cite{Aurell2011_vol106,Then2008,Schmiedl2007}, 
and specifically on phase transitions among optimal protocols \cite{Solon2018}, we identify similar phase transitions in optimal betting strategies.  We first prove such a result for uncorrelated races, and involving only two horses, which we then generalize to an arbitrary number of horses and to correlated races. In addition, we also give a general proof of the convexity in the most useful part of the front (positive part of the tradeoff branch), which rules out the existence of further phase transitions on that branch.

\section{Kelly's horse races}
\label{sec:def_model}

Let us recall here the main features of Kelly's horse race \cite{Kelly1956}. This race involves $M$ horses, which are numbered as $1,2...M$. 
The odds paid by the bookmaker when the horse $x$ wins is $o_x$, and the probability
for this to happen is $p_x$. A gambler can distribute his bets on the different horses, 
let $b_x$ be the fraction of the bet set on horse $x$, so that $\sum_{x=1}^M b_x=1$. 
For all $x$, $b_x>0$, because the gambler bets on all horses 
but only makes money from the horse $x$ that wins. 

A key feature of the model is that this dynamics is repeated, since all the money gained in one race 
is reinvested in the next race. Thus, 
the capital $C_{N+1}$ of the gambler after $N+1$ races is related to his capital after $N$ races, $C_N$,  by the expression
\be
C_{N+1}=o_xb_xC_{N}, \text{ with probability } p_x.
\label{eq_evol}
\ee
The important quantity is the long term growth rate of the capital 
which has the form
\be
\lim_{N \to \infty} \frac{1}{N} \ln C_N = \sum_x p_x \ln(o_x b_x),
\label{growth-rate}
\ee 
where the equality follows from the law of large numbers. 
Let us introduce the random variable $W_x=\ln(o_x b_x)$ which describes the contribution  
of horse $x$ to this growth rate. Its average with respect to the probability density $p_x$, 
is the long term growth rate denoted $\langle W \rangle$.

Kelly's strategy is defined from the optimization of this average growth rate over the betting strategy 
defined by $b_x$. A simple calculation given the constraint $\sum_x b_x=1$ leads to the proportional betting strategy $b_x^* =p_x$. This particular solution is independent of the odds $o_x$, but if there was a track take, the optimal solution would depend on both $o_x$ and $p_x$ \cite{Kelly1956}.

\begin{figure}
\centering
\includegraphics[scale=0.4]{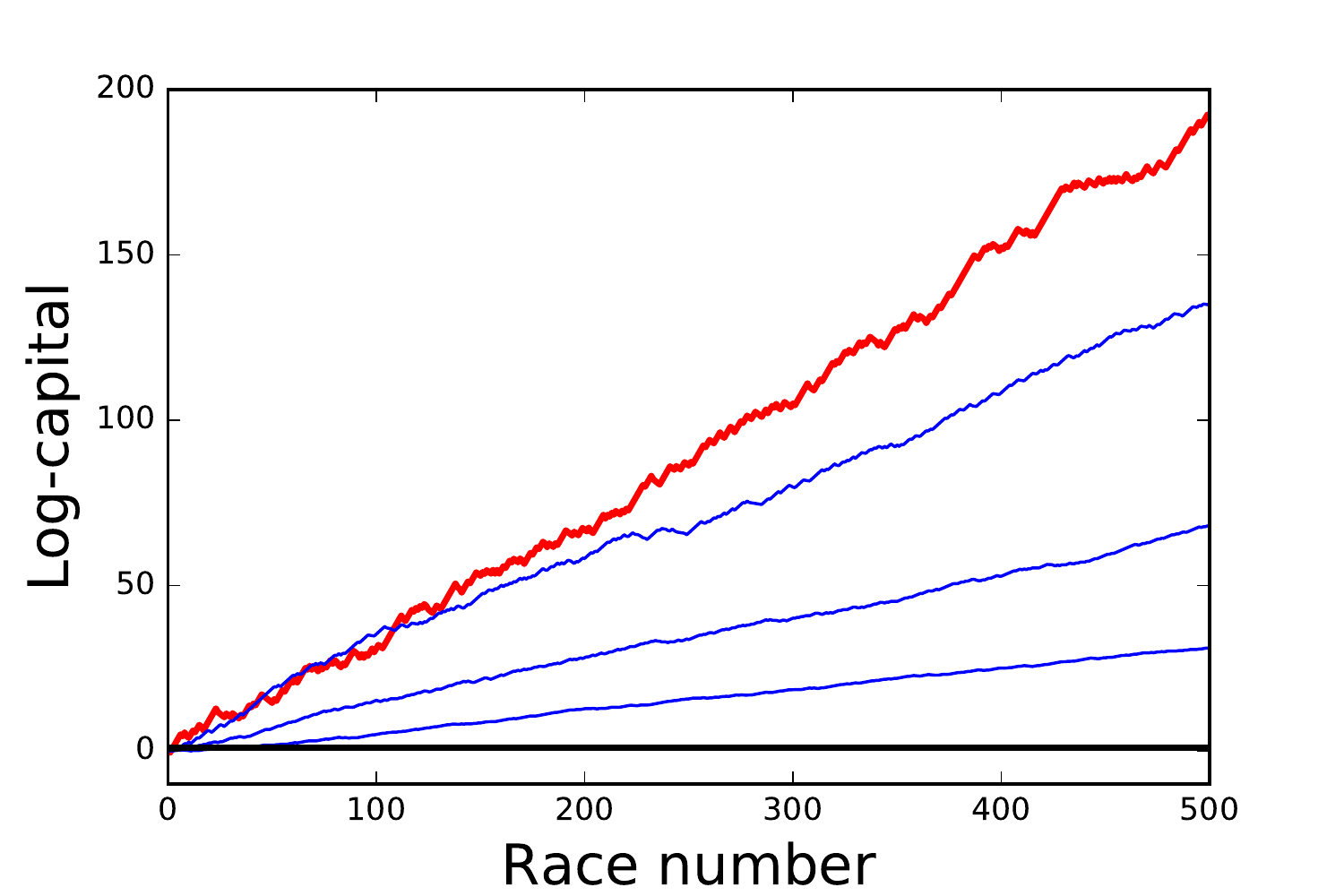} 
\caption{Logarithm of the capital of the gambler versus the number of races, for the optimal strategy (Kelly's) (thick red line) and for a selection of three non-optimal strategies (thin blue lines).} 
\label{fig1}
\end{figure}

Games of this type can be easily simulated in a computer using a random number generator to choose a winning horse for each race according to probability distribution $p_x$ and using Eq.~\eqref{eq_evol} to compute gambler's capital \cite{SM}. 
The growth of the capital is exponential and Kelly's strategy
dominates on long times all non-optimal strategies as shown in Fig. \ref{fig1}.

A central result of Stochastic Thermodynamics, namely fluctuation relations, 
can be obtained in a few steps for this model \cite{hirono_jarzynski-type_2015}.
Using the definition of $W$, and given that $b=p$ for Kelly's strategy, we obtain :
\be
\langle e^{-W} \rangle = \sum_x p_x \frac{1}{o_x p_x} = \sum_x \frac{1}{o_x} =1,
\label{JE}
\ee
where in the last equality, we have used the normalization of the distribution $r_x=1/o_x$ 
valid when there is no track take (fair odds). 
By Jensen's inequality, Eq. \eqref{JE} implies $\langle W \rangle \ge 0$, which also follows from 
 $\langle W \rangle = D(p | r) \ge 0$ where $D(p | r)$ denotes the Kullback-Leibler divergence between the distributions $p$ and $r$. 
This fluctuation relation (\ref{JE}) can be generalized for an arbitrary strategy of the gambler, not necessarily that of Kelly, and when the odds are not necessarily fair, by introducing the decomposition
$\tilde{W}_x=W_x + I_x,$
where $W_x=\ln(o_x b_x)$ as above, $\tilde{W}_x=\ln(o_x p_x)$ and $I_x=\ln(p_x /b_x)$. 
In this way, $\tilde{W}$ represents the growth rate of the gambler  
according to Kelly's strategy and $I$ 
measures the difference between the gambler's strategy and that of Kelly's in a KL sense, 
since $\langle I \rangle = D(p | b)$.
We have then
\be
\langle e^{-\tilde{W}} \rangle = \langle e^{-W -I} \rangle = \Lambda,
\label{SU}
\ee
with $\Lambda=\sum_x 1/o_x$.
In the same way that Eq.~\eqref{JE} is the analog of Jarzynski equality, Eq.~\eqref{SU} is similar to its generalization   
for absolutely irreversible processes \cite{Murashita2014}. By Jensen's inequality, the second-law like inequality
$\langle W \rangle \ge -\langle I \rangle - \ln \Lambda,$
follows which reduces to $\langle W \rangle \ge 0$ in the particular case of Kelly's strategy with fair odds. Note that in the general case, $\langle W \rangle$ can a priori be of any sign.

\section{Mean-variance tradeoff : Choice of utility function}
Kelly's strategy focuses on the maximization of the growth rate at the price of 
overlooking risk. Although bankruptcy is absent in Kelly's scenario because the growth of the capital is
geometric instead of arithmetic, the fluctuations of the capital are  
 large as shown in Fig. \ref{fig1} and represents a major concern.
This  problem has been widely recognized
in the gambling community. In practice gamblers and investors know that optimal Kelly can be “too risky”; and that “fractional Kelly” should be preferred, 
which deviates from the optimal solution but reduces the effective variance of the stochastic growth \cite{Thorp2011}. 
 
In the same spirit, we study here the optimal betting strategy that gives the highest capital growth rate while keeping a certain low value of risky fluctuations and analyze the corresponding trade-off between risk and gain. A similar idea is behind the mean-variance analysis introduced by Markowitz optimization \cite{Markowitz1952}. In contrast with Markowitz optimization however, which considers the mean and variance of the capital return in one race, we consider here the mean and the variance of the (long-term) growth rate of the capital after many races. This important conceptual difference allows us to recover Kelly's point as a special case of our analysis, whereas Kelly's point could not appear as a limiting case of Markowitz's optimization for this reason.
Hence, our utility function is    
a linear combination of the mean and standard deviation of the growth rate, namely $\langle W \rangle$ and $\sigma_W$ : 
\be
\label{J1}
\tilde{J}=\alpha \langle W \rangle - (1-\alpha) \sigma_W,
\ee
with $0 \le \alpha \le 1$.
In practice, we use the modified utility function 
\be
\label{objective}
J=\alpha \langle W \rangle - (1-\alpha) \sigma_W + \lambda \sum_x b_x,
\ee
where $\lambda$ is a Lagrange multiplier associated to the normalization of the bets.
An optimization of $J$ with respect to $b_x$ leads to   
$\lambda=-\alpha$. By reporting this into Eq.~\eqref{objective}, the optimal bets $b_x$ 
are solutions of : 
\be
\label{optimalb}
p_x-b_x = \frac{\gamma}{\sigma_W} p_x \left[ \ln(o_x b_x) - \langle W \rangle \right],
\ee
where $\gamma=(1-\alpha)/\alpha$.
 As expected, when $\alpha=1$ ($\gamma=0$), we recover the proportional betting of Kelly's strategy, which maximizes $\langle W \rangle$. Instead when $\alpha=0$ ($\gamma \to \infty$), we obtain the {\it null strategy} also called the risk free strategy, because in this case $\langle W \rangle=\sigma_W=0$. Between these two values, the strategy of the gambler is described as {\it  mixed} since it combines aspects associated to the optimization of $\langle W \rangle$ and $\sigma_W$.
  
\section{Exact solution for two horses}
Before embarking on the full problem with an arbitrary number of horses, it is instructive to analyze
the fully solvable case of two horses. 
Let the probability that the first horse wins (resp. loses) be $p$ (resp. $1-p$); the bet and the odd 
on the first (resp. second) horse are $b$ and $1/r$ (resp. $1-b$ and $1/(1-r)$) and let us introduce the parameter $\sigma=\sqrt{p(1-p)}$.

From the optimization of $J$, we obtain the 
optimal strategy $b^\pm$ :
\be
b^\pm=p \pm  \gamma \sigma,
\label{optimal bet}
\ee
where the $+$ (resp. $-$) sign 
corresponds to an overbetting (resp. underbetting) strategy with respect to 
Kelly's strategy where $b=p$.

\begin{figure}
\centering
\includegraphics[scale=0.45]{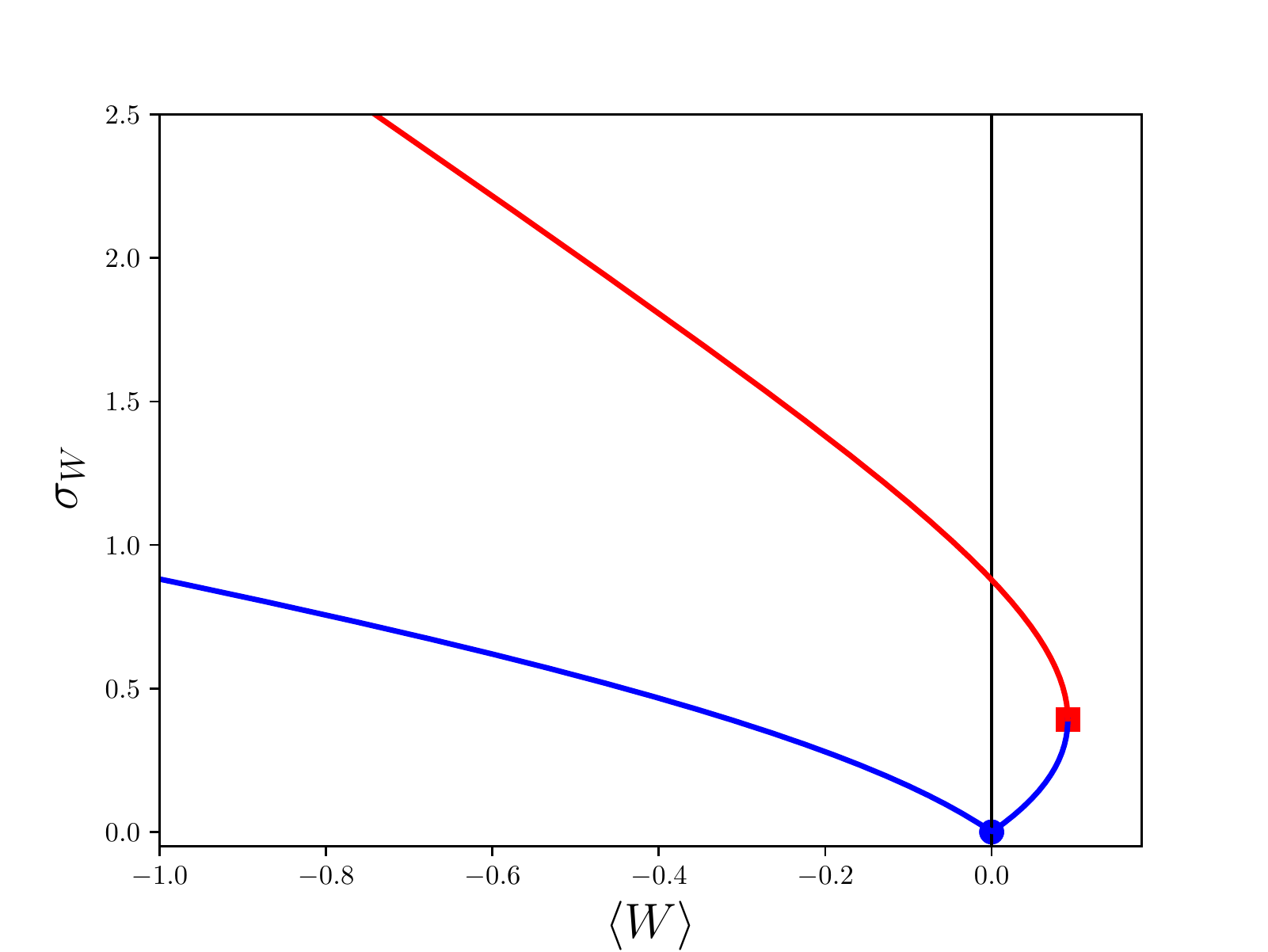} 
\caption{Trade-off branch (lower blue solid line) and non-trade-off branch (upper red solid line) in the plane ($\langle W \rangle$, $\sigma_W$) for two horses and for the parameters $(p=0.2,r=0.4)$. The two branches meet at the red square (Kelly's strategy), and the blue circle represents the null strategy. 
} 
\label{fig2}
\end{figure}

By reporting the optimal bet given by Eq.~\eqref{optimal bet} into the expression of $J$, one obtains the efficient border. 
As shown in Fig. \ref{fig2}, this border has two branches which meet at Kelly's point. 
When $p<r$ the lower blue solid line is the trade-off branch associated with $b^+$, 
while the upper red solid line is the non-trade-off branch, 
associated with $b^-$. The roles of $b^-$ and $b^+$ exchange when instead $p>r$. 
Let us first focus on the region where $\langle W \rangle \ge 0$.


We find that the slope of the Pareto border is 
\be
\left. \frac{d \sigma_W}{d \langle W \rangle} \right|_\gamma = \frac{\sigma}{p-b},
\ee
where $b$ is equal to $b^-$ when $r<p$ \cite{SM}. 
Therefore the slope of the Pareto border is infinite
at Kelly's point where $b^\pm=p$; 
while it reaches a finite value near the null strategy, namely
\be
\label{2gammac}
\left. \frac{d \sigma_W}{d \langle W \rangle} \right|_{\gamma_c} =\frac{1}{\gamma_c}=\frac{\sigma}{|p-r|}.
\ee
This signals a phase transition at this critical value $\gamma_c$, 
where the optimal strategy changes from the null strategy to a mixed strategy.
As a result, the optimal $J$ versus $\gamma$ changes from zero when $\gamma \ge \gamma_c$ (null strategy) to a non-zero value when $\gamma \le \gamma_c$ (mixed strategy). For two horses, such a plot is similar to what is shown for three horses in the inset of Fig. \ref{fig3}. 

To prove the existence of the phase transition, 
we have checked that the border is convex near the null strategy. 
It is indeed the case since 
\be
\label{2derivative}
\left. \frac{d^2 \sigma_W}{d \langle W \rangle^2} \right|_{\gamma =\gamma_c} = \frac{r(1-r)}{\sigma^2 \gamma_c^3} >0.
\ee
In the rest of this paper, we now focus on the general case for an arbitrary number of horses.

\section{Numerical results}
\begin{figure}
\includegraphics[scale=0.45]{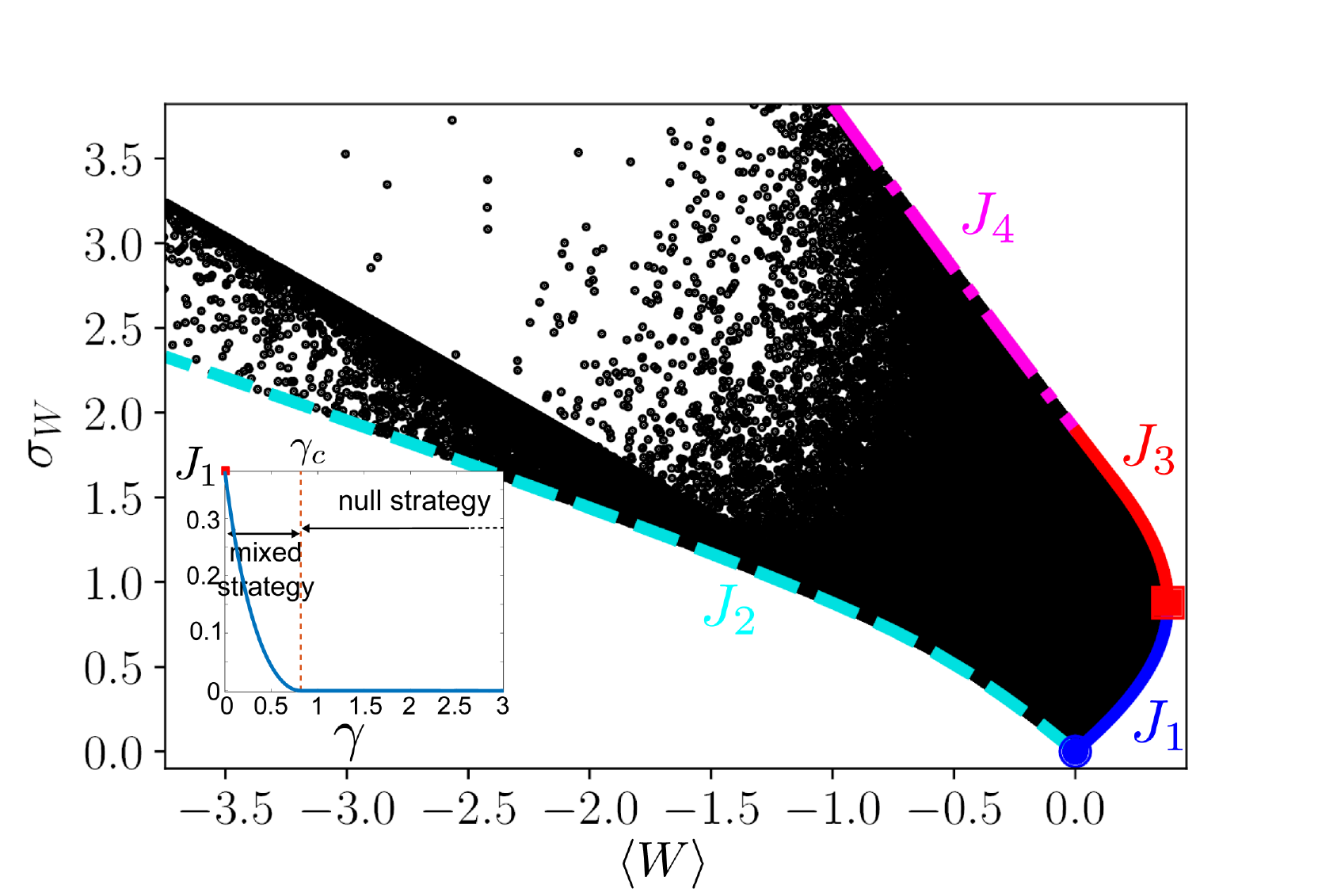}
\caption{Pareto borders for 3 horses obtained from numerical optimization of the utility functions $J_1,J_2,J_3$ and $J_4$ (colored solid lines), together with a cloud of points generated by randomly choosing bets satisfying all relevant constraints. Parameters are $p_1=0.2$, $p_2=0.6$, $r_1=0.4$ and $r_2=0.2$ for the first two horses. Inset: $J_1$ versus $\gamma$ along the trade-off branch ({\it i.e.} on the dark blue border).} 
\label{fig3}
\end{figure}

Let us now explain how to obtain the Pareto front from a numerical optimization of the utility function 
using a simulated annealing algorithm, as illustrated in Fig. \ref{fig3} for the case of three horses. 
Similarly to the case of two horses case, the lower and upper branch correspond to different optimization problems. The lower branch is formed by bets that maximize the growth rate $\langle W\rangle$ for a given value of the fluctuations $\sigma_W$, whereas the upper branch corresponds to maximal fluctuations $\sigma_W$ for a given value of the growth rate $\langle W\rangle$.

For the lower branch, there are two regions where $\langle W\rangle$ is either positive or negative. In the former case, the front is 
convex and can be recovered by the maximization of the utility function $J=J_1$ defined in Eq.~\eqref{J1}. 
In contrast, in the negative $\langle W\rangle$ region, the front is concave and a different strategy is needed. Following \cite{Solon2018}, we use a quadratic objective function
\begin{equation}
  J_2=-(\langle W\rangle-W_0)^2-k\sigma_W.
\end{equation}
We use a global minus sign in order to keep the same maximization procedure, although we wish in fact to minimize both the value of $\sigma_W$ and the distance to a target value $W_0$ for the growth rate.  
By varying the target value $W_0$ from 0 to a sufficiently negative value we can draw the negative lower branch. Parameter $k$ weighs the importance between the constraint of $\langle W\rangle$ being close to $W_0$ or minimizing the value of the fluctuations. We took $k=0.5$ although other moderate values would do. 

Similarly, the upper branch with positive $\langle W \rangle$ is concave and  
corresponds to the maximization of the objective function
\begin{equation}
	J_3=\alpha \langle W\rangle+(1-\alpha)\sigma_W,
\end{equation}
where the plus sign before $\sigma_W$ now ensures the maximization of the fluctuations in contrast 
with the lower branch case.
The upper branch with negative $\langle W \rangle$ 
appears almost straight for large negative values of $\langle W \rangle$. Thus, although $J_3$ could still be used there, 
further numerical precision can be achieved by using a modified objective function  
\begin{equation}
	J_4=-(\langle W\rangle-W_0)^2+k\sigma_W,
	\label{eq_j4}
\end{equation}
where again the plus sign in front of $\sigma_W$ corresponds to the maximization of fluctuations. 

General conclusions can also be obtained for this model near special points. Near Kelly's point, we find that the slope of the Pareto border is always vertical. This means that in practice if one is willing to pay a sacrifice a small amount of the average growth rate, one can lower the fluctuations significantly, thereby accessing "safer'' strategies 
such as the blue curves in Fig. \ref{fig1}.
Near the null strategy, we find a similar phase transition as in the two horses case, 
which we now analyze in more details.

\section{Mean-variance trade-off : bounds}
We recall that $r_x:=1/o_x$ and we assume a fair game for which $\sum_x r_x =1$. Then let $q_x:=r_x/p_x$, so that the first two moments of $q$ are $\langle q \rangle=1$ and $\sigma_q^2:=\langle q^2\rangle-\langle q\rangle^2=\langle q^2\rangle -1$. Let us focus on the branch of positive $\langle W \rangle$.
In this case, we find the following inequality,  
\be
\label{UR}
\sigma_W \ge \frac{\langle W \rangle}{\sigma_q},
\ee
which has a similar structure as thermodynamic uncertainty relations \cite{Horowitz2019,uffink1999}, and which captures a general trade-off between the mean and the variance of the growth rate.

The proof goes as follows : we
consider the quantity $\sigma_q^2 \sigma_W^2$, since $\sigma_q^2=\langle q^2 \rangle -1$,
we have using the Cauchy-Schwarz inequality
\bea
\sigma_q^2 \sigma_W^2 &=& \langle (q-1)^2 \rangle \langle
 ( W - \langle W \rangle)^2 \rangle,  \nonumber \\ 
& \ge & \langle (q-1)(W- \langle W \rangle) \rangle^2; \nonumber \\
& \ge & \left( \langle q W \rangle - \langle W \rangle \right)^2, \label{LR}
\eea
Now since $\langle qW \rangle = \sum_x r_x/b_x\log(b_x/r_x)= -D(r | b) \le 0$, then Eq.\eqref{UR} follows. 
This inequality is saturated when $b_x=r_x$, which corresponds to the null strategy.

Similar inequalities can be derived using instead other relevant Kullback-Leibler divergences, such as
$D(b | p)$ or $D(r | p)$. To exploit the first divergence, we introduce the ratio $s_x= b_x/p_x$ which is 
also a normalized probability distribution similar to $q$, with a second moment $\sigma_s^2$. Then, following the same steps, we obtain an inequality for the quantity $I$ introduced in Eq (3) :
\be
\label{URI}
\sigma_I \ge \frac{\langle I \rangle}{\sigma_s},
\ee
which is saturated when $b_x=p_x$, {\it i.e.} for Kelly's strategy.
To exploit the second divergence, we now use the quantity $\tilde{W}$, and we obtain 
the inequality
\be
\label{URW}
\sigma_{\tilde{W}} \ge \frac{\langle \tilde{W} \rangle}{\sigma_q},
\ee
which is saturated when $p_x=r_x$.
Note that Eqs. \ref{URI} and \ref{URW} represent new bounds which complement the 
inequalities $\langle I \rangle \ge 0$ and $\langle \tilde{W} \rangle \ge 0$ obtained previously.

\section{Phase transition in optimal strategies}
In order to prove that there are no tighter bounds of this type, 
we carry out a perturbation calculation near the null strategy using the vector $\epsilon_x$ 
\be
o_x b_x=\frac{b_x}{r_x}=1+\eps_x. \label{eq:eps} 
\ee
To ensure that $\vec{b}$ is still a probability measure, we require that the column vector $\vec{\eps}=(\eps_x)_x$ lies on the hyperplane $(\vec{r},\vec{\eps})=\sum_x r_x \eps_x=0$. 

By evaluating $\langle W \rangle$ and $\sigma_W$ to first order
in $\vec{\eps}$, we find that 
%
$\sigma_W \sim \langle W \rangle/\gamma_c$, with
\BEQ 
\label{gammac}
\gamma_c = \sigma_q, 
\EEQ  
an expression which we can be checked by plotting a zoom of the Pareto border near the null strategy \cite{SM}.
The evaluation of the second order derivative at the null strategy on the Pareto border requires a calculation to second order in $\vec{\eps}$, which gives  
\be
\label{2derivative-gen}
\left. \frac{d^2 \sigma_W}{d \langle W \rangle^2} \right|_{\gamma =\gamma_c} = \frac{C}{ \gamma_c^5},
\ee
where 
$C=\langle q^3\rangle -\langle q^2\rangle^2$ \cite{SM}. By Cauchy-Schwarz again, it follows that 
$\langle q^2\rangle^2 = \langle q^{3/2} q^{1/2} \rangle^2 \le \langle q^3 \rangle$, thus $C\ge 0$, with equality iff $p_x=r_x$.

In the particular case of two horses, it is straightforward to check that the expression of $\gamma_c$ given in Eq. \eqref{2gammac} and that of the second derivative in Eq. \eqref{2derivative}
are recovered from Eqs.~\eqref{gammac}-\eqref{2derivative-gen}.
These calculations show that there is always a phase transition in this model near the null strategy for an arbitrary number of horses in the region of positive $\langle W \rangle$. 
A similar calculation shows that the slope has the opposite value on the other side in the region of negative $\langle W \rangle$. 

\section{Shape of the front : general results}
\subsection{Large negative growth rate}
In the regions of the phase diagram corresponding to negative values of $\langle W \rangle$, 
 the Pareto front is open. 
Namely, the growth rate diverges because it is evaluated on some $b_x\to 0$. 
Easy computations shows that points in the $(\langle W\rangle,\sigma_W)$ plane satisfy asymptotically 
$\langle W\rangle\to -\infty$ and
$\sigma_W / \langle W\rangle \to -\sqrt{(1-P')/P'}$ when
bets $b_{x'} \to 0$ for $x' \in X'$ with
$P':=\sum_{x'\in X'} p_{x'}$. 
The smallest slope (lower front), 
is obtained by putting all the bets on the horse  $x^*$ which has the least chances to win; this is the worst strategy.

\subsection{Lower front : positive growth rate}
In order to decide whether other phase transitions are possible in this model, we now study 
the convexity of the front near any point. More precisely, we define the front 
 as the extremum locus of the functional 
\bea
& \tilde{J}_{m^*}(b;\lambda,\mu) :=
 \langle W^2\rangle 
 + \lambda(\langle W\rangle -m^*) \nonumber \\ 
 &+ 2\mu (\sum_x b_x-1),
\eea
where $\lambda,\mu$ are Lagrange multipliers fixing $\langle W\rangle$ and implementing the bet normalization constraint. The procedure is equivalent to extremizing the variance for a given
average value $m^*$. 
The null gradient condition $D \tilde{J}_{m^*}(b;\lambda,\mu)=0$ defines $(b,\lambda,\mu)$ as an implicit
function $f(m^*)$ of $m^*$.
The gradient of $f$, which is the Hessian of $\tilde{J}_{m^*}$, may be inverted with some efforts, yielding by the implicit function theorem
the slope $d\sigma_W/d\langle W\rangle=d\sigma_W/dm^*$ and then finally, the
second derivative $d^2\sigma_W/d\langle W\rangle^2$ in terms of $\mu$
(proportional to the inverse of the Pareto slope parameter $\gamma$)
and averaged
functionals of $b_x/p_x$. Explicit formulas given in Supp. Mat \cite{SM} have been checked numerically. 
One can then prove in whole generality that the part of the lower front between the null strategy and Kelly's strategy is convex, turning to concave in some neighborhood  of the null strategy when
$\langle W\rangle <0$, and some neighborhood of Kelly's strategy on the upper front, as confirmed numerically in Fig. \ref{fig2} and Fig. \ref{fig3} in the case of two and three horses. 
Note that this calculation does not exclude the possibility of other phase transitions in other parts of the front.

\section{Correlated races}
As a variation on Kelly's horse races, we now assume that the races are no longer independent but follow from an ergodic Markov process defined by the conditional probability $p_{x|y}$, which represents the probability that the horse $x$ wins if the previous horse that won the race was horse $y$. Let the bets be also conditional and defined by $b_{x|y}$ such that $\sum_x b_{x|y}=1$. The odds denoted by $o_x=1/r_x$ are assumed to be fair $\sum_x r_x=1$. 
The average growth rate $\langle W \rangle$ now takes the following form  
\be
\langle W \rangle= \lim_{N \to \infty} \langle W_N \rangle = \sum_{x,y} p_{x|y} \bar{p}_y \ln(b_{x|y} o_x),
\ee
where $\bar{p}_y$ denote the unique steady state probability of the races.
By optimizing $\langle W \rangle$ with respect to $b_{x|y}$, we find that
the optimal strategy is still proportional betting with now $p_{x|y}=b_{x|y}$.
This is the new Kelly's strategy for this case.

\begin{figure}
\includegraphics[scale=0.45]{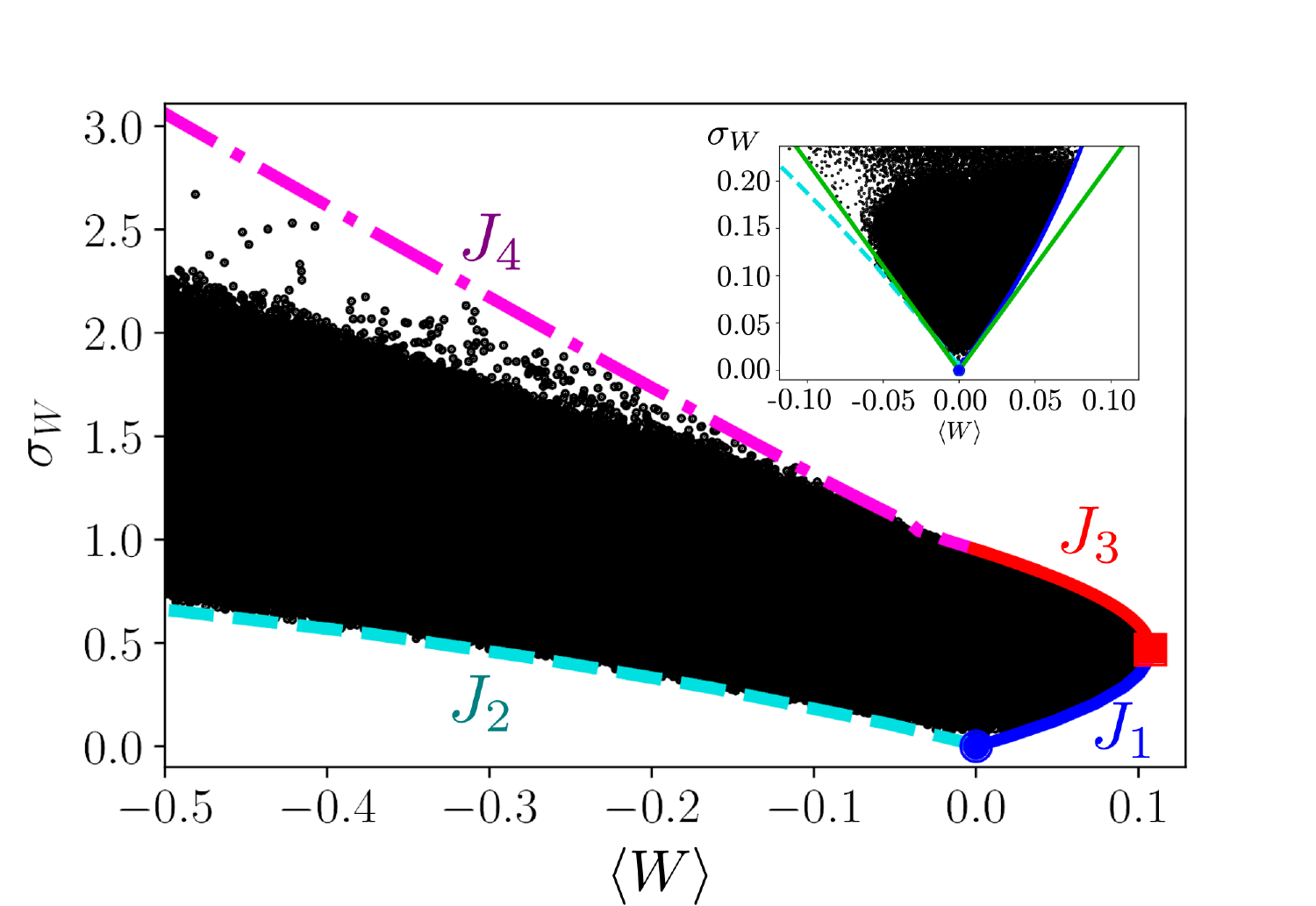} 
\caption{Same plot as in fig. \ref{fig3} but for the case of for 3 horses in the presence of correlations between the races. Parameters are detailed in Ref. \cite{SM}. Inset: zoom near the null strategy together with predictions from linear approximation.} 
\label{fig4}
\end{figure}

On the trade-off branch, the relevant utility function is 
\be
\label{objective_corr}
J=\alpha \langle W \rangle - (1-\alpha) \sigma_W + \sum_y \lambda_y \sum_x b_{x|y},
\ee
where $\lambda_y$ are Lagrange multipliers associated to the normalization of the bets.
The Pareto borders are shown in Fig. \ref{fig4}.
We observe numerically that when correlations are present the upper front for negative W becomes convex in some intermediate region. In that region, the border can not longer be described by $J_3$ and the use of $J_4$ is unavoidable.


The null strategy corresponds to the condition that for any $x,y$, $b_{x|y}=r_x$, in which case both 
the average growth rate and its variance are zero. An expansion with respect to that strategy can be carried as before.
The $q$ distribution is now defined as $q_{x|y}=r_x/p_{x|y}$, which is a probability distribution because 
\be
\langle q \rangle = \sum_{xy} p_{x|y} \bar{p}_y \frac{r_x}{p_{x|y}}=\sum_{xy} \bar{p}_y r_x=1.
\ee
Its second moment is now $\langle q^2 \rangle=\sum_{xy} p_{x|y} \bar{p}_y q_{x|y}^2$. Except for this modification, the critical $\gamma$ takes the same form as in Eq. \eqref{gammac}, which
is numerically tested in the inset of Fig. \ref{fig4}.

An inequality similar to Eq. \eqref{UR} can also be obtained in the case of correlated races because in this case 
the conditional bets $b_{x|y}$ are still a probability distribution $\sum_x b_{x|y}=1$, and therefore 
following the same steps, the positivity of $D(r | b)$ leads to a similar result.
In fact, the normalization of $q$ is equivalent to a fluctuation relation generalizing Eq. \eqref{SU}\cite{hirono_jarzynski-type_2015}. 
because
in that case 
\be
\langle e^{-W} \rangle = \sum_{xy} p_{x|y} \bar{p}_y \frac{1}{p_{x|y} o_x}=1,
\ee
while
\be
\langle e^{-W-I} \rangle = \Lambda,
\ee
holds in the general case for an arbitrary strategy with $I_{x|y}=\ln(p_{x|y} /b_{x|y})$. 

\section{Conclusion}
In this work, we have derived general Fluctuation Relations for betting models of Kelly's type, and 
a bound on the average capital growth rate, similar to 
thermodynamic uncertainty relations. This bound  
captures the classic trade-off between average growth rate 
and risk, which plays a central role in money investment \cite{Bouchaud2009}. 
In models with repetitive investment dynamics,
all utility functions become under suitable conditions equivalent to a utility function 
with a log mean variance form \cite{Thorp2011}, 
 which is the form considered here. 
This suggests that our work should be applicable to a broad class of econophysics models,
for which log utility functions are used.   

In our work, we have identified a phase transition between the null strategy and a mixed strategy, 
and we have shown that there is no other phase
transition between the null strategy and Kelly's point due to the convexity of the lower front. 
We have also illustrated how to handle non-convex utility functions, an important issue 
for applications to machine learning \cite{bottou2016optimization}. 

The explicit analytical expressions which we have obtained for the slope and curvature of the front at any point could be used to move directly along the front, as an alternative to the involved optimization algorithm used here.
It would be also interesting to explore more systematically how additional constraints affect the efficient border. The question of adaptative optimization of the bets, where possible non-Markovian or non-ergodic features could arise, is a rich  inference problem worth pursuing \cite{tal_adaptive_2020}.  
Finally, we hope that this framework could open news research directions on evolutionary trade-offs and Pareto optimality in biology \cite{Shoval2012,Seoane2015}. 

\acknowledgments
L.D. acknowledges financial support from Spanish Ministerio de Economía, Industria y Competitividad through grant FIS2017-83709-R.
We acknowledge many insightful discussions with L. Peliti and E. Aurell.

\bibliographystyle{apsrev4-2}
\bibliography{./ref2.bib}


\begin{widetext}

\vskip 0.8 cm

\appendix{\centerline{\large\sf Supplementary Material}}

\vskip 0.3 cm


\section{Notations.} 
Let $r_x:=1/o_x$,   $W_x:=\ln(o_x b_x)=\ln(b_x/r_x)$ , $q_x:=\frac{r_x}{p_x}$.
Denote by $\langle \ \cdot\ \rangle$ the average of a function $f=(f_x)_{x}$ with respect to the weights $(p_x)_x$,
$\langle f\rangle:=\sum_x p_x f_x$. Then the average growth rate is
 \BEQ \langle W \rangle=\sum_x p_x \ln(b_x/r_x), \EEQ
  	and the standard deviation of a given
strategy $\vec{b}=(b_x)_x$ is 
\BEA \sigma_W&:=&\sqrt{\langle W\rangle^2-\langle W\rangle^2}
\nonumber\\
&=& \sqrt{\sum_x p_x \ln^2(b_x/r_x)-\langle W\rangle^2}.
\EEA

\Medskip By hypothesis, the column vectors $\vec{p}=(p_x)_x$, $\vec{r}=(r_x)_x$  and $\vec{b}=(b_x)_x$ are probability distributions,
$\sum_x p_x=\sum_x r_x=\sum_x b_x=1$.  Furthermore,
 the two first moments of $q$ are
\BEQ \langle q\rangle=1,\qquad  \sigma_q^2:=\langle q^2\rangle-
\langle q\rangle^2=\langle q^2\rangle -1.\EEQ


\section{Solution for two horses}


Let the probability that the first horse wins (resp. loses) be $p$ (resp. $1-p$); the bet and the odd 
on the first (resp. second) horse are $b$ and $1/r$ (resp. $1-b$ and $1/(1-r)$) and let us introduce the parameter $\sigma=\sqrt{p(1-p)}$.
In this way, the odds are fair. 
Let also $\gamma:=\frac{1-\alpha}{\alpha}$ and $\sigma:=\sqrt{p(1-p)}$. Then

\BEQ 
\label{W}
\langle W \rangle=p\ln(\frac{b}{r})+(1-p)\ln(\frac{1-b}{1-r}),
\EEQ
and
\BEQ 
\label{SW}
\sigma_W^2=p(1-p) \ln^2 \frac{b(1-r)}{(1-b)r} 
= \left(\sigma \ln \frac{b(1-r)}{(1-b)r} \right)^2.  \EEQ

From the optimization of the utility function $J$ defined in the main text, 
we obtain the optimal strategy $b^\pm$ :
\be
b^\pm=p \pm  \gamma \sigma,
\label{optimal bet}
\ee
where the $+$ (resp. $-$) sign corresponds to an overbetting (resp. underbetting) strategy with respect to Kelly's strategy where $b=p$. As shown in Fig. 2 of main text, these two solutions form the two branches of the efficient border which meet at Kelly's point. 
When $p<r$ the lower blue solid line is the tradeoff branch associated with $b^+$, 
while the upper red solid line is the non-tradeoff branch, 
associated with $b^-$. The roles of $b^-$ and $b^+$ exchange when instead $p>r$. 
Let us first focus on the region where $\langle W \rangle \ge 0$ and let us assume e.g. $p>r$, in which case 
\BEQ b=p-\gamma\sigma. \EEQ

Using Eqs. \ref{W}-\ref{SW}, we find
\Bigskip
\BEQ \frac{d\langle W \rangle}{db}=\frac{p-b}{b(1-b)} =
\frac{\gamma\sigma}{b(1-b)},  \label{f1} \EEQ and
\BEQ \half \frac{d (\sigma_W^2)}{db}=\frac{\sigma^2}{b(1-b)} 
\ln (\frac{b(1-r)}{(1-b)r}). \label{f2} \EEQ
Hence by taking the ratio of Eq. (\ref{f2}) and (\ref{f1}), 
\BEQ \half \frac{d( \sigma_W^2)}{d\langle W \rangle}=
\frac{\sigma}{\gamma} \ln (\frac{b(1-r)}{(1-b)r}) = \frac{\sigma_W}{\gamma}>0. \label{f3} \EEQ

Using the definition of $\gamma$, we deduce that the slope of the Pareto border is 
\be
\left. \frac{d \sigma_W}{d \langle W \rangle} \right|_\gamma = \frac{\sigma}{p-b},
\ee
where $b$ is equal to $b^-$ since we have assumed $r<p$. 
This equation shows that the slope becomes infinite
at Kelly's point where $\gamma \to 0$ and $b^- \to p$; 
while it reaches a finite value near the null strategy, namely
\be
\label{gammac}
\left. \frac{d \sigma_W}{d \langle W \rangle} \right|_{\gamma_c} =\frac{1}{\gamma_c}=\frac{\sigma}{p-r}.
\ee
This suggests that there is a phase transition between the null strategy and a mixed strategy at this critical value $\gamma_c$. To confirm this point, we need to check that the border is convex near the null strategy. 

To do so, we take the derivative of Eq. \ref{f3} with respect to $\langle W \rangle$ as before : 
\BEQ \half \frac{d^2 (\sigma_W^2)}{d\langle W\rangle^2}=  \frac{\frac{\sigma}{\gamma} 
\frac{d}{db}\ln(\frac{b}{1-b})  + \frac{\sigma_W}{\sigma\gamma^2}}{d\langle W\rangle/db} = \frac{1}{\gamma^2}+ b(1-b) \frac{\sigma_W}{\sigma^2\gamma^3}; 
\label{f4}
\EEQ

Finally, using the general formula
\BEQ (\sqrt{f})''=\frac{1}{2\sqrt{f}} f''-\frac{1}{4} f^{-3/2}
(f')^2, \label{eq:sqrt''}
\EEQ we find the simple result
\BEQ \frac{d^2 \sigma_W}{d\langle W\rangle^2}=   
\frac{b(1-b)}{\sigma^2\gamma^3},
\EEQ
which is always positive, in particular near the null strategy where it takes the value  
\be
\label{2derivative}
\left. \frac{d^2 \sigma_W}{d \langle W \rangle ^2} \right|_{\gamma =\gamma_c} = \frac{r(1-r)}{\sigma^2 \gamma_c^3} >0.
\ee


\section{General expansion near the null strategy to first order}

Let us now analyze the general case for an arbitrary number of horses.
As already observed for the two horses case, the lower and upper branch correspond to different optimization problems. The lower branch is formed by bets that maximize the growth rate $\langle W\rangle$ with the minimal average fluctuations $\sigma_W$ whereas the upper branch corresponds to maximal fluctuations $\sigma_W$ for a given value of the growth rate $\langle W\rangle$.
For the lower branch, there are two regions where $\langle W\rangle$ is either positive or negative. In the former case, the front is 
convex and can be recovered by the maximization of the utility function $J=J_1$ defined in the main text. 

General conclusions can be obtained for this model near special points. Near Kelly's point, we find that the slope of the Pareto border is always vertical. 
To prove this, we rely on perturbation calculations near these specific strategies. 
In this case, we find that the first order correction 
to $\langle W \rangle$ vanishes, while that of $\sigma_W$ does not vanish.
It follows from this that the slope of the border $(\langle W \rangle, \sigma_W)$ is indeed vertical near Kelly's point.

We now detail the expansion near the null strategy, where we find 
a similar phase transition as found in the two horses case. 
Let us introduce the vector $\epsilon_x$ to measure the distance to the null strategy as
\be
o_x b_x=\frac{b_x}{r_x}=1+\eps_x. \label{eq:eps} 
\ee
To ensure that $\vec{b}$ is still a probability measure, we require that the column vector $\vec{\eps}=(\eps_x)_x$ lies on the hyperplane $(\vec{r},\vec{\eps})=\sum_x r_x \eps_x=0$. 

By evaluating $\langle W \rangle$ and $\sigma_W$, we find that 
\BEQ \langle W \rangle \sim \langle \eps\rangle- \half \langle 
\eps^2 \rangle , \label{eq:leading-<w>} \EEQ
and
\BEQ \sigma_W^2 \sim \langle \eps^2\rangle-\langle \eps\rangle^2.  \label{eq:leading-sigmaw2}
\EEQ
To leading order in $\eps$ or in $\sigma_W$, the equation for the optimal bets, namely 
Eq. (8) of the main text yields
\BEQ \eps_x-\langle\eps\rangle \sim  \frac{\sigma_W}{\gamma} (1-q_x)  \label{eq:David-b*-ter}
\EEQ
Multiplying (\ref{eq:David-b*-ter}) by $r_x$ and summing over $x$ to eliminate $\eps_x$ yields
\BEQ \langle \eps\rangle\sim \frac{\sigma_W}{\gamma}(\langle 
q^2\rangle -1)  \label{eq:David-b*-qua}
\EEQ
But we can also use Eq. \ref{eq:David-b*-ter} to obtain to leading order in $\eps$, 
\BEQ \langle \eps^2\rangle-\langle\eps\rangle^2 
\sim  \frac{\sigma_W^2}{\gamma^2} \sigma_q^2.
\EEQ
Combining these two equations, we obtain the slope on the tradeoff branch near the null strategy
\BEQ
\label{slope}
\sigma_W \sim \frac{\langle W \rangle}{\gamma_c} = \frac{\langle W \rangle}{\sigma_q}. 
\EEQ


\section{General expansion near the null strategy to second order}

Before embarking on the evaluation of the second order derivative near the null strategy on the Pareto border, it is useful to formalize the general problem of minimization of the variance, 
not necessarily near the null strategy. In general the minimization of the variance leads to the optimal bet $b^*=b(m)$ parametrized by a given value of the average, $\langle W \rangle=m$. 
Below, we focus on the case where $m \ge 0$.
The following  parametrization of the bets is then appropriate to explore the space of parameters $b$ around $b(m)$ :
 \BEQ \frac{b_x}{r_x}=\frac{b^*_x}{r_x}+\eps_x, \EEQ
coinciding with
(\ref{eq:eps}) when $b^*=r$ is the null strategy. 
To simplify the notation, we will drop the subscript with the star on the $b_x$ and on $m$, since it will be implicit that we consider this optimal solution.

Taking 
into account the constraints $\langle W \rangle=m$ and  $(\vec{r},\vec{\eps})=0$, this
is equivalent to minimizing the functional $-\tilde{J}_{m}$ (equivalent to maximizing $\tilde{J}_{m})$
 \BEQ -\tilde{J}_{m}(\eps;\lambda,\mu):=\langle W^2 \rangle + \lambda (\langle W \rangle -
m) + 2\mu (\vec{r},\vec{\eps}). \EEQ 
Thus, we require 
the null gradient condition $\tilde{\nabla} \tilde{J}_{m}=0$, 
where $\tilde{\nabla}:= \left(\begin{array}{c} \nabla \\
\partial_{\lambda} \\ \partial_{\mu} \end{array}\right)$, and
$\nabla=(\partial_{\eps_x})_x$.
 
The general formula for minimization with constraints may be
found in standard textbooks, implying positivity of  ${\cal H}=-\nabla^2 \tilde{J}_{m}=\nabla^2 \langle W^2\rangle + \lambda \nabla^2 (\langle W \rangle -m)+2\mu \nabla^2 (\vec{r},\vec{\eps})$. 
 Computations yields for the null gradient condition,
\BEQ -\half \nabla \tilde{J}_{m}= \Big\{ \ln(b_x/r_x) +\frac{\lambda}{2} \Big\} 
\frac{r_x}{b_x}p_x + \mu r_x=0. \label{eq:nablaJm}
\EEQ

After multiplying Eq. (\ref{eq:nablaJm}) by $b_x/r_x$ and summing over $_x$, one obtains
\BEQ -\frac{\lambda}{2}=m
+ \mu  \label{eq:lambdaJm}
\EEQ
Thus $b=b(m)$ is given by two conditions:
\BEQ \begin{cases} \ln \left( \frac{b_x}{r_x} \right )+ \mu \left( \frac{b_x}{p_x}-1 \right)=m \\ \sum_x b_x=1 \end{cases}  \label{eq:b*}
\EEQ
The r.-h. s. of the first line of (\ref{eq:b*}), a constant, is fixed  by averaging, yielding the trivial relation 
$\langle W \rangle=m$, whence the need for an extra condition given by the second line, which
defines a function $\mu(m)$. 
By comparing Eq. \ref{eq:b*} with the equation for the optimal bets given in Eq. 8 of the main text, we obtain
the general expression : 
\BEQ
\label{mu}
\mu=\frac{\sigma_W}{\gamma}.
\EEQ
On the Pareto front, the differential of $\tilde{J}$ defined in Eq. 6 of the main text must vanish. This leads to 
the condition $d\langle W \rangle - \gamma d \sigma_W=0$, which means that $d\sigma_W/dm=1/\gamma$. Then, using Eq. \ref{mu}, we obtain the equally general result
\BEQ
\label{mu-gen}
\mu=\half \frac{d \sigma_W^2}{dm}.
\EEQ

Let us now focus on the expansion near the null strategy. We already now from
 the expression for the slope of the border derived in the previous section, namely Eq. \ref{slope}, 
that $\frac{d\sigma_W}{dm}\to_{m\to 0}\frac{1}{\sigma_q}$.
Let us further assume that $\frac{d^2 \sigma_W}{dm^2}\to_{m\to 0} \tilde{C}$ 
where $\tilde{C}$ or equivalently $C$ given by
$C\equiv \sigma_q^5 \tilde{C}$ is an unknown coefficient to be determined self-consistently
which controls the curvature of the border near the null strategy.

Now using Eq. \ref{mu-gen}, \BEQ \frac{d\mu}{dm}= \half \frac{d^2 \sigma_W^2}{dm^2} =
(\frac{d\sigma_W}{dm})^2 + \sigma_W \frac{d^2\sigma_W}{dm^2}
\sim (\frac{1}{\sigma_q}+ m\tilde{C})^2 + \tilde{C}m/\sigma_q
\sim \frac{1}{\sigma_q^2}+ 3\frac{\tilde{C}}{\sigma_q}m. \EEQ
Since to dominant order, $m\sim \sigma_q^2 \mu$, we have
\BEQ
m\sim \sigma_q^2 \mu (1-\frac{3}{2}\tilde{C}\sigma_q^3 \mu), 
\EEQ

We now evaluate $\eps_x$ to second order, 
using  $b_x/r_x=1+\eps_x$ in 
Eq. \ref{eq:b*}. This yields :
\BEQ \eps_x-\frac{\eps_x^2}{2} + \mu [ \frac{r_x}{p_x}(1+\eps_x)-1]
\sim  \sigma_q^2 \mu (1-\frac{3}{2}\tilde{C}\sigma_q^3 \mu). 
\EEQ
Thus
 $\eps_x\sim (1+\sigma_q^2-\frac{r_x}{p_x})\mu+O(\mu^2)$ to dominant order, which gives 
\BEA \eps_x &\sim& (1+\sigma_q^2-\frac{r_x}{p_x})\mu  +\frac{\eps_x^2}{2}-\mu \frac{r_x}{p_x}\eps_x - \frac{3}{2} \tilde{C}\sigma_q^5 \mu^2 \nonumber\\
&\sim & (1+\sigma_q^2-\frac{r_x}{p_x})\mu  + \Big\{ 
\half(1+\sigma_q^2 - \frac{r_x}{p_x}) (1+\sigma_q^2 - 3\frac{r_x}{p_x}) - \frac{3}{2} C\Big\} \mu^2 +O(\mu^3).
\EEA

This expansion must be consistent with the condition $\sum_x r_x\eps_x=0$. 
Using the above formula, one finds that the first order term vanishes, and the second 
order term vanishes only if and only if 
\BEQ \half\langle  \ (1+\sigma_q^2 -\frac{r_x}{p_x})(1+\sigma_q^2 -
3\frac{r_x}{p_x}) \frac{r_x}{p_x} \ \rangle = \frac{3}{2} C,
\EEQ
which means that 
$3C=(1+\sigma_q^2)^2 - 4 (1+\sigma_q^2) \langle q^2\rangle +3\langle q^3\rangle$, or 
\BEQ C = \langle q^3\rangle - (1+\sigma_q^2)^2 = \langle q^3\rangle -
(\langle q^2\rangle)^2. \qquad (C) \EEQ

Therefore, we have shown that the second derivative is
\be
\label{2derivative-gen}
\left. \frac{d^2 \sigma_W}{d^2 \langle W \rangle} \right|_{\gamma =\gamma_c} = \frac{C}{ \gamma_c^5}.
\ee
Note that the positivity of second derivative is guaranteed in the general. Indeed 
using Cauchy-Schwarz,  
$\langle q^2\rangle^2 \le \langle q^3 \rangle$, which shows that $C\ge 0$, with equality iff $r_x^{1/2}\propto \frac{r_x^{3/2}}{p_x}$, i.e. when $p_x=r_x$.

In the end, we obtain the following second order approximation of the Pareto border near the null strategy
\be
\label{2ndexpansion}
\sigma_W \sim \frac{1}{\gamma_c} \langle W \rangle + \frac{C}{2 \gamma_c^5} \langle W \rangle^2.
\ee

In the particular case of two horses, it is straightforward to check that the expression of $\gamma_c$ given in Eq. \ref{gammac} and that of the second derivative in Eq. \ref{2derivative}
are recovered from Eq.~\ref{gammac} and Eq.~\ref{2derivative-gen} because of the following relations :
\BEQ
\sigma_q^2=\frac{r^2}{p}+\frac{(1-r)^2}{1-p}-1=\frac{(r-p)^2}{\sigma^2},
\EEQ
and
\BEA C&=&-\sigma_q^4+\langle (q-1)^3\rangle +\sigma_q^2 =-\frac{(p-r)^4}{\sigma^4} + \frac{(p-r)^3}{\sigma^4} (2p-1) + \frac{(p-r)^2}{\sigma^2} \nonumber\\ 
&=& \frac{(p-r)^2}{\sigma^4} \Big\{ -(p-r)^2+(p-r)(2p-1)+p(1-p)\Big\} \nonumber\\
&=& \frac{(p-r)^2}{\sigma^4} r(1-r),
\EEA
therefore
\BEQ
\frac{C}{\gamma_c^5}=\frac{r (1-r)}{\gamma_c^3 \sigma^2},
\EEQ as expected.


\section{General mathematical results for the shape of the border}


We prove in this section the general results stated p.4 and 5 of Main text.

\Bigskip {\bf V. 1. Large negative values of growth rate.}

\Medskip Let $\emptyset \subsetneq X'\subsetneq X$. Assume  $b_x\to 0$ when $x\in X'$, while all other parameters $(b_x)_{x\not\in X'}$ are kept fixed.   Eq. (\ref{eq:b*}),
\BEQ \ln(\frac{b_x}{r_x})+ \mu \frac{b_x}{p_x}=m+\mu \qquad {\mathrm{independent\ from}}\ x
\EEQ
 implies 
that $\ln(b_x/r_x)\sim m+\mu$ independently of $x$ for all $x\in X'$; thus the speed of convergence of  $b_x$ to $0$ for  $x\in X'$ may be characterized
by a single parameter $\eps\to 0^+$ defined by  $-1/\eps=\ln(b_{x'}/r_{x'})$, $x'$ being 
an arbitrary element of $X'$; for all $x\in X'$, $\ln(b_x/r_x)\sim -1/\eps$.  
Hence a first relation, where $0<P':=\sum_{x\in X'} p_x<1$,
$$ (1)\qquad m= \sum_x p_x \ln(b_x/r_x) \sim \sum_{x\in X'} p_x \ln(b_x/r_x) \sim_{\eps\to 0} -P'/\eps $$
Now, $-1/\eps \sim m+\mu$, hence a second relation,
$$ (2)\qquad m\sim_{\eps\to 0} P'(m+\mu).$$
From (1) and (2), we deduce $\mu \sim_{\eps\to 0} -(1-P')/\eps$. 

\Medskip Similarly,
\BEQ \sigma_W^2=\sum_x p_x \ln^2(b_x/r_x) \ - \ m^2 \sim P'/\eps^2 - m^2= P'(1-P')/\eps^2
\EEQ
whence a $P'$-dependent asymptote for the front, 
\BEQ \frac{\sigma_W}{|m|} \longrightarrow_{\eps\to 0} \sqrt{\frac{1-P'}{P'}}.
\EEQ
This is in whole coherence with the two-horse case analyzed p.3.
This gives a set of asymptotes with slopes $\Big\{ \sqrt{\frac{1-P'(X')}{P'(X')}} \Big\}$, where $P'(X'):=\sum_{x\in X'} p_x$, and  $X'$ ranges in the set of non trivial subsets of $X$. Note that the function $P'\mapsto \sqrt{\frac{1-P'}{P'}}$ is decreasing, so the largest slope (highest asymptote) is $\sqrt{\frac{1-p_{min}}{p_{min}}}$, with $p_{min}:=\min_{x\in X} p_x$ (say, $p_{min}=p_{x^*}$ for some $x^*\in X$), while the smallest slope (lowest asymptote) is the inverse quantity, $\sqrt{\frac{p_{min}}{1-p_{min}}}$, obtained by choosing
$X'=X\setminus\{x^*\}$  -- clearly, the worst possible strategy, since all bets are set on the worst horse -- . 

\bigskip \noindent
{\bf V. 2. Convexity of the lower front.} The remainder of the section is devoted to the computation of the second derivative of the border $\frac{d^2 \sigma_W}{d\langle W\rangle^2}$.  We introduce the functional 
\BEQ \tilde{J}_{m^*}(b;\lambda,\mu):=\langle W^2\rangle + \lambda(\langle W\rangle -m^*) + 2\mu (\sum_x b_x-1),
\EEQ
 where $\lambda,\mu$ are Lagrange multipliers fixing $\langle W\rangle$ and implementing the bet normalization constraint. 

\Medskip Conditions $\frac{\partial \tilde{J}_{m^*}}{\partial\lambda}=0$, 
$\frac{\partial \tilde{J}_{m^*}}{\partial\mu}=0$ fix the 
average growth rate $\langle W\rangle$ to the value $m^*$, and
impose the constraint $\sum_x b_x=1$. Conditions $\frac{\partial \tilde{J}_{m^*}}{\partial b_x}=0$ then give the local extrema
of $\sigma_W$ for fixed $\langle W\rangle$ since $\sigma_W^2=\langle W^2\rangle - (m^*)^2$. Depending on the eigenvalues of the Hessian
$D^2 \tilde{J}_{m^*}$, one may in principle select local maxima or minma; in practice this is however complicated due to the constraints.

\begin{Theorem} \label{th:second-derivative}
Let us introduce the vectors
\BEQ 
a_x:=\half (\nabla^2 \tilde{J}_{m})_{xx}=p_x(\frac{r_x}{b_x})^2 \Big( 1-\frac{\lambda}{2} - \ln(b_x/r_x)\Big)
\EEQ
and  $u_x=p_x \frac{r_x}{b_x}$ and $v_x=2r_x$
so that 
$a_x=\frac{u_x}{p_x}(u_x+ \frac{\mu}{2} v_x).$ 
Then, 
  \BEA &&\frac{d^2 \sigma_W}{d\langle W\rangle^2}=
\frac{4\mu}{ \sigma_W^2 
 {\mathrm{det}}^2_a(u;v)} \times \nonumber\\
&& \qquad \times \Big\{
\frac{\sigma_W^2}{\mu^2} \Big[ (
1-\half(u,v)_a)-\frac{\mu}{4} 
{\mathrm{det}}^2_a(u;v) \Big] - \frac{\mu}{4}  {\mathrm{det}}_a^2(u;v)  \Big\} \EEA

where $(\cdot,\cdot)_a$, $\det^2_a(\cdot,\cdot)$ refer to the
(not necessarily positive-definite) "pseudo-metric" $g_{xy}=a_x^{-1}\del_{x,y}$ on $\R^{|X|}$, namely,
\BEQ (u,v)_a=\sum_x a_x^{-1} u_x v_x, \qquad  |u|^2_a=(u,u)_a,  |v|^2_a=(v,v)_a, \qquad {\mathrm{det}}^2_a(u,v)=|u|^2_a |v|^2_a - (u,v)_a^2. 
\EEQ

\end{Theorem}

The above analytical formula is implicit, since 
$(b,\lambda,\mu)$ are functions of $\langle W\rangle$. 
Furthermore, it implies the following result:

\begin{Theorem} \label{th:convexity} 
\begin{enumerate}

\item[(i)] The part of the  {\em lower front} (see section 3) -- i.e. of
the variance-minimizing curve -- with $\langle W\rangle\, >0$ is
strictly convex, i.e. $\frac{d^2\sigma_W}{dm^2}>0$.
\item[(ii)] In some neighborhood of the null strategy on the left lower front defined by $\langle W\rangle <0$, and in some neighborhood of Kelly's strategy on the upper front,  the front is concave, i.e. satisfies $\frac{d^2\sigma_W}{dm^2}<0$.
\end{enumerate}
\end{Theorem}

\bigskip\noindent
{\bf V. 2. 1. Proof of Theorem \ref{th:second-derivative}.} 

Let us start from  $(b=(b_x)_{x\in X}=b^*,\lambda=\lambda^*,\mu=\mu^*;m=m^*)$ such that $f(b,\lambda,\mu;m)\equiv D\tilde{J}_{m^*}=0$.  The gradient of $f$ w.r. to $(b,\lambda,\mu)$, which is the Hessian $D^2 \tilde{J}_m$, is (as proved
below) invertible. The implicit function theorem then implies that
the locus $\{f=0\}$ is given locally around $(b^*,\lambda^*,\mu^*;m^*)$ by functions $b=b(m), \lambda=\lambda(m), \mu=\mu(m)$ such that 
\BEQ \frac{db_x}{dm}=-\Big((\tilde{\nabla} f)^{-1}  \frac{\partial f}{\partial m} \Big)_x,  \qquad
 \frac{d\lambda}{dm}=-\Big((\tilde{\nabla} f)^{-1}  \frac{\partial f}{\partial m} \Big)_{\lambda}, \qquad \frac{d\mu}{dm}=-\Big((\tilde{\nabla} f)^{-1}  \frac{\partial f}{\partial m} \Big)_{\mu}  \label{eq:dblambdamu/dm}
\EEQ
In subsequent computations, we use rescaled variables $\eps_x=
\frac{b_x-b^*_x}{r_x}$ instead of $b$.  The notation $(..)_x$ denotes the component along $\eps=(\eps_x)_{x\in X}$ of the gradient. 
The condition $\sum_x \frac{b_x}{r_x} \nabla \tilde{J}_{m}=0$
 yields $-\frac{\lambda}{2}=m^*+\mu$, which implies in turn
the equations
\BEQ \ln(b_x/r_x)+\mu(\frac{b_x}{p_x}-1)=m, \EEQ
which are equivalent to (6) (see Main text) if one sets $\mu=\sigma_W/\gamma$. 

\Medskip
{\em A. 1. Computation of the inverse of the Hessian.} \BEQ \tilde{\nabla}f=\left( \begin{tabular}{c|cc} $\nabla^2 \tilde{J}_{m} $ & $\partial_{\lambda}{\nabla} \tilde{J}$ & 
$\partial_{\mu}{\nabla} \tilde{J}$
 \\ \hline
$\Big(\partial_{\lambda}{\nabla} \tilde{J}\Big)^t$ & 
$\partial_{\lambda}^2 \tilde{J}_{m}$ & $\partial^2_{\lambda\mu}
\tilde{J}_{m}$ \\ 
$\Big(\partial_{\mu}{\nabla} \tilde{J}\Big)^t$ & 
$\partial_{\lambda\mu}^2 \tilde{J}_{m}$ &$ \partial^2_{\mu}
\tilde{J}_{m}$
\end{tabular} \right)   =  
\left( \begin{tabular}{c|c} $A $ & $B$
 \\ \hline
$B^t$ & 
$0$
\end{tabular} \right) 
\EEQ

where
\BEQ A=\nabla^2 \tilde{J}_{m}=(\nabla^2_{xy} \tilde{J}_{m})_{xy}, \qquad B=(\partial_{\lambda}{\nabla} \tilde{J} \ \ 
\partial_{\mu}{\nabla} \tilde{J}) \EEQ
are $|X|\times |X|$, resp. $|X|\times 2$ matrices, and 
$A=A^t$ is symmetric. The inverse of the matrix
$\tilde{\nabla}f$ is, as follows from a simple computation,
\BEQ (\tilde{\nabla}f)^{-1}=  
\left( \begin{tabular}{c|c}  $A^{-1}\Big(I-B (A^{-1}BC^{-1})^t\Big)$ & $A^{-1}BC^{-1}$ \\ \hline $\Big((A^{-1})BC^{-1}\Big)^t$ & $-C^{-1}$
\end{tabular} \right) 
\EEQ
where $C=B^t A^{-1} B$ is a symmetric $2\times 2$ matrix. 
 Some
elementary algebra yields in block form

{\small \BEQ (\tilde{\nabla}f)^{-1}= \frac{1}{\det^2_a(u;v)}  \left(
\begin{tabular}{c|cc} $ \half(\det^2_a(u;v) \ {\mathrm{diag}}(a^{-1})- Z) $ & $ |v|^2_a \, \overrightarrow{a^{-1}u} - (u,v)_a  \overrightarrow{a^{-1}v}$ & $|u|^2_a \ \overrightarrow{a^{-1}v} - (u,v)_a \ \overrightarrow{a^{-1}u} $
\\ \hline $ |v|^2_a \,  \big(a^{-1}u\big)^t - (u,v)_a  \big(a^{-1}v)^t$  & $-2|v|^2_a$ & $2(u,v)_a$ \\ 
$ |u|^2_a \,  \big(a^{-1}v\big)^t - (u,v)_a  \big(a^{-1}u)^t$  & $2(u,v)_a$ & $-2|u|^2_a$
\end{tabular} \right)  \label{eq:nablafinverse}
\EEQ }

Note that $u_x\sim a_x\sim p_x$ in the neighborhood of the null
strategy, so that (rescaling vectors $c,d\in\R^{|X|}$ by the 
probability weights $p_x$,  $(pc)_x:=p_x c_x, (pd)_x:=p_x d_x$)
$(pc,pd)_a \sim \sum_x p_x c_x d_x$ boils down to the usual
$L^2$-metric weighted by $p$ in that limit.   
In the equation above, we have denoted $Z$ the  following symmetric 
  $|X|\times |X|$-matrix,
 \BEQ Z:=  |v|^2_a \   (a^{-1}u \otimes 
a^{-1}u) + |u|^2_a \ (a^{-1} v\otimes a^{-1}v) - (u,v)_a \ 
\Big\{ (a^{-1}u\otimes a^{-1}v)+(a^{-1}v\otimes a^{-1}u) \Big\}
\label{eq:W}
\EEQ
Then $\overrightarrow{a^{-1}u}$ is a vector in $\R^{|X|}$, $(a^{-1}u)_x=a_x^{-1} u_x$, with transpose $(a^{-1}u)^t$.

\Bigskip {\em A. 2.} We easily derive from the inverse Hessian formula the {\em first derivative} of the Pareto border. 
We find first
\BEQ \half \frac{d  \sigma^2_{W} }{dm}=\half \frac{d
 \langle W^2\rangle}{dm} - m
=\half \sum_x \frac{d\eps_x}{dm} \frac{\partial \langle W^2\rangle}{\partial\eps_x} \ - \ m \label{eq:dsigmaw2/dm}\\
\EEQ
since $\langle W^2\rangle$ is independent of $\lambda,\mu$ and does not depend explicitly on $m$
\BEA
&&\qquad\qquad\qquad =  \sum_x p_x \frac{r_x}{b_x}  \ln(b_x/r_x)  \ 
\Big(-(\tilde{\nabla} f)^{-1}  \frac{\partial f}{\partial m} \Big)_x
 -m  =  \sum_x p_x \frac{r_x}{b_x}  \ln(b_x/r_x)  \ 
\Big((\tilde{\nabla} f)^{-1}  \left(\begin{array}{c} 0 \\ 1 \\ 0 \end{array}\right) \Big)_x
 -m. \nonumber\\
  \label{eq:first-derivative}
\EEA
Then 
\BEA && \half \frac{d\sigma_W^2}{dm}=-m+  \sum_x p_x \frac{r_x}{b_x}  \ln(b_x/r_x)  \ 
\Big((\tilde{\nabla} f)^{-1}  \left(\begin{array}{c} 0 \\ 1 \\ 0 \end{array}\right) \Big)_x \nonumber\\
&& =-m+\frac{1}{{\mathrm{det}}^2_a(u;v)}  \sum_x \Big(mu_x-\frac{\mu}{2}
(v_x-2u_x)\Big) \Big(|v|_a^2 \frac{u_x}{a_x}-(u,v)_a \frac{v_x}{a_x} \Big)\nonumber\\
&&= \frac{1}{{\mathrm{det}}^2_a(u;v)} \mu \Big( |v|_a^2 
\sum_x \frac{u_x}{a_x} - (2-\frac{\mu}{2}|v|_a^2) \sum_x 
\frac{v_x}{a_x} \Big) \nonumber\\
&&= \mu \label{eq:58}
\EEA 
The result is coherent with the slope formula $d\sigma_W/d
\langle W\rangle=1/\gamma$, if one sets $\mu=\sigma_W/\gamma$. 

\Medskip {\em A. 3. Exploration of the front: a possible application of the inverse Hessian formula.} Formula (\ref{eq:dblambdamu/dm}), completed by the expression (\ref{eq:nablafinverse}) for $(\tilde{\nabla}f)^{-1}$, allows a numerical exploration of the 
front starting from an arbitrary point. This provides an elementary  alternative
to the numerical exploration process described in Main Text.

\Medskip {\em B. 1. Second derivative. Preliminary results.} We must differentiate once
more to obtain $d^2 \sigma_W^2/d\langle W\rangle^2=d^2 \sigma_W^2/dm^2$. This is straightforward in principle using the explicit expression for $(\tilde{\nabla}f)^{-1}$, but in practice, computations are rather involved. 

\Medskip Differentiating  (\ref{eq:dsigmaw2/dm}) once again, one finds
\BEQ 1+\half \frac{d^2 (\sigma_W^2)}{dm^2}   \equiv 
D^2_1+D^2_2+D^2_3, 
\EEQ
with
\BEQ D^2_1:=\half  \sum_x   \left(\frac{d\eps_x}{dm}\right)^2 \frac{\partial^2
(\langle w^2\rangle)}{\partial \eps_x^2}
=\sum_x a^0_x \Big((\tilde{\nabla} f)^{-1}   \left(\begin{array}{c} 0 \\ 1 \\ 0 \end{array}\right)  \Big)_x^2 
\EEQ
where 
\BEQ a^0_x:=\half  p_x \frac{d^2}{d\eps^2} \ln^2(\frac{b_x}{r_x}+\eps) \, \Big|_{\eps=0}=
p_x  (\frac{r_x}{b_x})^2 (1-\ln(b_x/r_x))=\frac{u_x}{p_x} \Big\{ (1-m-\mu)u_x+\frac{\mu}{2}
v_x \Big\} ; \label{eq:a}
\EEQ

\BEA && D^2_2:=\half 
\frac{d\lambda}{dm} \ \sum_x \frac{\partial \langle w\rangle^2}{\partial \eps_x} \ \frac{\partial}{\partial\lambda} \left(
\frac{d\eps_x}{dm}\right) \nonumber\\
&&\qquad =\Big((\tilde{\nabla} f)^{-1}  \left(\begin{array}{c} 0 \\ 1 \\ 0 \end{array}\right) \Big)_{\lambda} \ \sum_x  p_x 
\frac{r_x}{b_x} \ln(\frac{b_x}{r_x}) \ \frac{\partial}{\partial\lambda} \Big((\tilde{\nabla} f)^{-1}  \left(\begin{array}{c} 0 \\ 1 \\ 0 \end{array}\right) \Big)_{x}
\EEA

\BEA && D^2_3:=
\half \sum_{x,y} \frac{\partial
\langle  w^2\rangle}{\partial\eps_x}\  \frac{d\eps_y}{dm} \ 
\frac{\partial}{\partial\eps_y} \left(\frac{d\eps_x}{dm}\right) 
\nonumber\\
&& \qquad = \sum_x p_x \frac{r_x}{b_x} \ln(\frac{b_x}{r_x})
\ \sum_y 
 \Big((\tilde{\nabla} f)^{-1}  \left(\begin{array}{c} 0 \\ 1 \\ 0 \end{array}\right) \Big)_{y}\ 
 \frac{\partial}{\partial\eps_y}  \Big((\tilde{\nabla} f)^{-1}  \left(\begin{array}{c} 0 \\ 1 \\ 0 \end{array}\right) \Big)_{x}.
  \label{eq:second-derivative}
\EEA

\Medskip Upon computing these quantities for $(\eps,\lambda,\mu;m)
=(0,\lambda(m),\mu(m);m)$, we can use (\ref{eq:b*}) and
replace in the sum $\ln(b_x/r_x)$ by $m-\mu(\frac{b_x}{p_x}-1)$.

\Medskip The only coefficients of the inverse Hessian that we need are
\BEQ  \Big((\tilde{\nabla} f)^{-1}  \ \left(
\begin{array}{c} 0 \\ 1 \\ 0 \end{array}\right)  \Big)_x= 
\frac{1}{\det^2_a(u;v)} \Big(  |v|^2_a \, \frac{u_x}{a_x} - 
(u,v)_a\, \frac{v_x}{a_x} \Big).  \label{eq:only-coeff-need}
\EEQ

and

\BEQ  \Big((\tilde{\nabla} f)^{-1}  \ \left(
\begin{array}{c} 0 \\ 1 \\ 0 \end{array}\right)  \Big)_{\lambda}=-\frac{2|v|_a^2}{\det^2_a(u;v)}.
\EEQ

\Bigskip {\em B. 2. Remarkable identities} Computations for the  second derivative involve
the following sums,  $I_{i,j}=\sum_x a_x^{-1}\, u_x^i v_x^j$
 $(i+j=2)$, $I_{i,j}=\sum_x a_x^{-1}\,  \frac{u_x^i v_x^j}{p_x a_x}$ 
$(i+j=4)$, 
$I_{i,j}=\sum_x a_x^{-1} \, \frac{u_x^i v_x^j}{(p_x a_x)^2}$
 $(i+j=6)$. 
The explicit expressions for $a$ in terms of $u,v$, together with the normalization condition $\sum_x p_x=\sum_x b_x=1,$ yield a set
of {\em remarkable identities} giving a triangular structure,  
$$I_{4,0}+\frac{\mu}{2} I_{3,1}=|u|_a^2, \qquad 
\frac{\mu}{2} I_{2,2}+I_{3,1}=(u,v)_a,  \qquad I_{2,2}+\frac{\mu}{2} I_{1,3}= |v|_a^2$$
and
$$  I_{6,0}+\frac{\mu}{2}I_{5,1}=I_{4,0}, \qquad 
I_{5,1}+\frac{\mu}{2} I_{4,2}=I_{3,1}, 
$$
$$  I_{4,2}+\frac{\mu}{2} I_{3,3}=I_{2,2}, \qquad
I_{3,3}+\frac{\mu}{2} I_{2,4}=I_{1,3}
$$
The more elementary identities
\BEQ |u|_a^2+\frac{\mu}{2} (u,v)_a=1,\qquad \frac{\mu}{2}
|v|_a^2 + (u,v)_a=2
\EEQ
can be proven similarly, from which
\BEQ {\mathrm{det}}^2_a(u;v)=(1-\frac{\mu}{2}(u,v)_a)\frac{2}{\mu}
(2-(u,v)_a)-(u,v)_a^2 = \frac{2}{\mu}(2-(\mu+1)(u,v)_a).
\label{eq:det2auv}
\EEQ

They imply after tedious computations:
\BEA && D^2_1=  \frac{1}{{\mathrm{det}}_a^4(u;v)} 
\Big\{ (1-m-\mu) \Big[ - \frac{8}{\mu} I_{3,1} + \frac{4}{\mu^2} \Big(4- (4+2\mu) (u,v)_a
+(1+2\mu) (u,v)_a^2 \Big)\Big] \nonumber\\
&&\qquad\qquad + \frac{\mu}{2} \Big[- \frac{8}{\mu} I_{2,2}   - \frac{8}{\mu^2} (u,v)_a
((u,v)_a-2) \Big] \Big\}.
\EEA

\Medskip {\em B. 3. (computation of  the second term $D^2_2$)}.  {\em We prove that $D^2_2=0$.}    Let 
$\tilde{M}(\lambda):=\tilde{\nabla}f=$ \\ $\Big($
\begin{tabular}{c|cc} $ M(\lambda)$ & Cst
\\ \hline Cst  & Cst
\end{tabular}$ \Big) $;
then
\BEQ\frac{\partial}{\partial\lambda} \Big((\tilde{\nabla} f)^{-1}  \left(\begin{array}{c} 0 \\ 1 \\ 0 \end{array}\right) \Big)_{x}
=-\Big(\tilde{M}(\lambda) \frac{d\tilde{M}}{d\lambda} \tilde{M}^{-1}(\lambda) \left(\begin{array}{c} 0 \\ 1 \\ 0 \end{array}
\right) \Big)_x 
\EEQ
\qquad $= \Big( (\tilde{\nabla}f)^{-1}  \ \Big($  \begin{tabular}{c|cc} $ {\mathrm{diag}}(p_x(\frac{r_x}{b_x})^2)$ & 0
\\ \hline 0  & 0
\end{tabular}$ \Big) \ (\tilde{\nabla}f)^{-1} \left(\begin{array}{c} 0 \\ 1 \\ 0 \end{array}
\right) \Big)_x  $

\qquad $ = \frac{1}{{\mathrm{det}}^2_a(u;v)} \ 
\Big( (\tilde{\nabla} f)^{-1}  \ \Big($  \begin{tabular}{c|cc} $ {\mathrm{diag}}(\frac{u^2_x}{p_x})$ & 0
\\ \hline 0  & 0
\end{tabular}$ \Big) \  \left(\begin{array}{c} 
|v|_a^2 \frac{u_x}{a_x}-(u,v)_a \frac{v_x}{a_x} \\ -2|v|_a^2 \\ 
2(u,v)_a \end{array}
\right) \Big)_x  $

\BEQ =  \frac{1}{{\mathrm{det}}^2_a(u;v)} \ \Big( (\tilde{\nabla}f)^{-1} \  \left(\begin{array}{c}  w_x \\ 0 \\ 0 \end{array}\right)
\Big)_x =  \frac{1}{2 {\mathrm{det}}^4_a(u;v)} \ 
 \Big( ({\mathrm{det}}^2_a(u;v) {\mathrm{diag}}(a^{-1})-Z)w \Big)_x,
 \EEQ
 where using the  above remarkable identities,
 \BEQ w_x:=\frac{u^2_x}{p_x} (|v|_a^2 \frac{u_x}{a_x} -(u,v)_a
 \frac{v_x}{a_x})=\frac{u_x^2}{p_x} ( \frac{4}{\mu} \frac{u_x}{a_x} - (u,v)_a \frac{v_x+\frac{2}{\mu}u_x}{a_x})=
 \frac{1}{p_x a_x} \Big\{ \frac{4}{\mu} u_x^3 - (u,v)_a (
 \frac{2}{\mu} u_x^3+u_x^2 v_x)\Big\}.
 \EEQ

 \Medskip {\em Computation of $(Zw)$-term.} $Z \equiv Z_1+Z_2-(Z_3+Z_3^{sym})$, with
$Z_1:=|v|_a^2 (a^{-1} u\otimes a^{-1} u)$, $Z_2:=|u|_a^2
(a^{-1}v\otimes a^{-1}v)$, $Z_3:=(u,v)_a (a^{-1}u\otimes a^{-1}v)$, $Z_3^{sym}:=(u,v)_a (a^{-1} v\otimes a^{-1}u)$. 
Using the remarkable identities, we get

\BEQ (Z_1w)_x=|v|_a^2 \frac{u_x}{a_x} \Big\{ \big(\frac{4}{\mu}-\frac{2}{\mu}(u,v)_a\big) I_{4,0} -(u,v)_a I_{3,1} \Big\} 
=|v|_a^2 \frac{u_x}{a_x} \Big\{ \frac{2}{\mu}|u|_a^2 (2-(u,v)_a)
-2I_{3,1}\Big\}  \label{eq:W1}
\EEQ
\BEQ (Z_3w)_x= (u,v)_a \frac{u_x}{a_x} \Big\{ \frac{2}{\mu}
(2-(u,v)_a)I_{3,1}-(u,v)_a I_{2,2} \Big\} = 
(u,v)_a \frac{u_x}{a_x} \frac{2}{\mu} \Big\{ 2I_{3,1}-(u,v)_a^2
\Big\}   \label{eq:W3}
\EEQ
\BEA && ((Z_1-Z_3)w)_x=\frac{u_x}{a_x} \Big[ I_{3,1} \Big\{-2|v|_a^2 -
\frac{4}{\mu} (u,v)_a\Big\} + (\frac{2}{\mu})^2
(1-\frac{\mu}{2} (u,v)_a) (2-(u,v)_a)^2 + \frac{2}{\mu} (u,v)_a^3\Big]  \nonumber\\
&&= \frac{u_x}{a_x} \Big\{ -\frac{8}{\mu}
I_{3,1}    +  \frac{4}{\mu^2} \Big[4 - (4+2\mu) (u,v)_a
+(1+2\mu) (u,v)_a^2 \Big]  \Big\}  \label{eq:W1-W3}
\EEA

\Medskip Similarly,
$(Z_2 w)_x=|u|_a^2 \frac{v_x}{a_x}  \frac{2}{\mu} \Big\{ 2I_{3,1}-(u,v)_a^2
\Big\} $
(compare with  (\ref{eq:W3}));
$(Z_3^{sym} w)_x= (u,v)_a \frac{v_x}{a_x}  \Big\{ \frac{2}{\mu}|u|_a^2 (2-(u,v)_a) -2I_{3,1}\Big\} 
$
(compare with (\ref{eq:W1}));
$
((Z_2-Z_3^{sym})w)_x=\frac{v_x}{a_x} \Big\{ \frac{4}{\mu} I_{3,1}   - \frac{4}{\mu} |u|_a^2 (u,v)_a\Big\}. $  
The contribution of the term in $(Z_1-Z_3)w$ to 
$C_2:=2{\mathrm{det}}^4_a(u;v) \sum_x p_x \frac{r_x}{b_x} \ln(\frac{b_x}{r_x}) \ \frac{\partial}{\partial\lambda} \Big((\tilde{\nabla} f)^{-1}  \left(\begin{array}{c} 0 \\ 1 \\ 0 \end{array}\right) \Big)_{x}$ is
\BEA && \sum_x (mu_x-\frac{\mu}{2}(v_x-2u_x)) \frac{u_x}{a_x}
\Big\{ -\frac{8}{\mu} I_{3,1} + \frac{4}{\mu^2} [4-(4+2\mu) 
(u,v)_a+(1+2\mu) \, (u,v)_a^2 ] \Big\} \nonumber\\
&&= (m+\mu) |u|_a^2 \Big\{-\frac{8}{\mu} I_{3,1} + 
\frac{4}{\mu^2} [4-(4+2\mu) \, (u,v)_a + (1+2\mu) \, (u,v)_a^2] \Big\}\nonumber\\
&&\qquad -\frac{\mu}{2} (u,v)_a  \Big\{ -\frac{8}{\mu} I_{3,1}
+ \frac{4}{\mu^2} [4-(4+2\mu) \, (u,v)_a + (1+2\mu) \, (u,v)_a^2] \Big\} \label{eq:C2-13}
\EEA

Similarly, the contribution of the term in $(Z_2-Z_{3,sym})w$
 to $C_2$ is
\BEA && \sum_x (mu_x-\frac{\mu}{2}(v_x-2u_x)) \frac{v_x}{a_x}
\Big\{ \frac{4}{\mu} I_{3,1} - \frac{4}{\mu} |u|_a^2 (u,v)_a
\Big\} \nonumber\\
&&= (m+\mu) (u,v)_a \Big\{  \frac{4}{\mu} I_{3,1} - \frac{4}{\mu} |u|_a^2 (u,v)_a
\Big\}  -\frac{\mu}{2} |v|_a^2  \Big\{  \frac{4}{\mu} I_{3,1} - \frac{4}{\mu} |u|_a^2 (u,v)_a
\Big\}   \label{eq:C2-23sym}
\EEA

\Bigskip
{\em Contribution of the diagonal term.} 
Finally, the contribution of the term in ${\mathrm{det}}^2_a(u;v) \frac{w_x}{a_x}$ to $C_2$ is 
\BEA && {\mathrm{det}}^2_a(u;v) \sum_x (mu_x-\frac{\mu}{2}(v_x-2u_x)) \frac{1}{p_x a_x^2} \Big\{ \frac{4}{\mu} u_x^3 -
(u,v)_a (\frac{2}{\mu}u_x^3+u_x^2v_x)\Big\} \nonumber\\
&&= {\mathrm{det}}^2_a(u;v)  \Big\{ (m+\mu) \big[ \frac{4}{\mu} (|u|_a^2 -\frac{\mu}{2} I_{3,1})-\frac{2}{\mu}|u|_a^2 (u,v)_a \big] \nonumber\\
&&\qquad - \frac{\mu}{2} \big[ \frac{4}{\mu} I_{3,1} - \frac{2}{\mu} (u,v)_a^2 \big] \Big\}  \label{eq:C2diag}
\EEA

Adding up the three terms (\ref{eq:C2-13}), (\ref{eq:C2-23sym}) and (\ref{eq:C2diag}), and using formula (\ref{eq:det2auv}) for
${\mathrm{det}}^2_a(u;v) $, one gets zero after some
extra work. 

The partial derivative of $(\tilde{\nabla}f)^{-1}$ w.r. to
$\eps_y$ is computed as (compare with {\em  B.3.})  $-(\tilde{\nabla}f)^{-1} \frac{\partial \tilde{\nabla} f}{\partial\eps_y} (\tilde{\nabla}f)^{-1}$. Now,

\BEQ (\frac{\partial}{\partial\eps_y}A)_{xx}=
\frac{\partial}{\partial\eps_y} \Big[ \frac{2p_x}{(\frac{b_x}{r_x}+\eps_x)^2}(1-\frac{\lambda}{2}-\ln(\frac{b_x}{r_x}+\eps_x))
\Big] \Big|_{\eps=0}=-2\del_{x,y} p_y(\frac{r_y}{b_y})^3 (3-\lambda-2\ln (\frac{b_y}{r_y}))
\EEQ

Writing $B=(B_{\cdot,1}\ B_{\cdot,2})$, with  $(B_{\cdot,1})_x=
(\frac{p_x}{\frac{b_x}{r_x}+\eps_x})$, $(B_{\cdot,2})=(2r_x)_x$,
we get $\nabla B_{\cdot,2}=0$ and 
\BEQ (\frac{\partial}{\partial\eps_y} B_{\cdot,1})_x\Big|_{\eps=0}=
-\del_{x,y} p_y (\frac{r_y}{b_y})^2
\EEQ

All together, all coefficients of $\frac{\partial}{\partial\eps_y}
\tilde{\nabla}f\Big|_{\eps=0}$ vanish, except three of them,
\BEQ  \alpha_y:=\Big(\frac{\partial}{\partial\eps_y}\tilde{\nabla}f\Big)_{y,y}=-2p_y(\frac{r_y}{b_y})^3 (3-\lambda-2\ln (\frac{b_y}{r_y}))=-4 (\frac{u_y}{p_y})^2 (\frac{3}{2}u_y+\frac{\mu}{2} v_y)
\label{eq:alphay}
\EEQ
and
\BEQ \beta_y:=\Big(\frac{\partial}{\partial\eps_y}\tilde{\nabla}f\Big)_{\lambda,y}=\Big(\frac{\partial}{\partial\eps_y}\tilde{\nabla}f\Big)_{y,\lambda}= - p_y (\frac{r_y}{b_y})^2
=-\frac{u_y^2}{p_y}.  \label{eq:betay}
\EEQ

Then, successively,

\begin{itemize}
\item[(i)]  the vector $w_y:=\Big(\frac{\partial}{\partial\eps_y}
\tilde{\nabla}f\Big) \ (\tilde{\nabla}f)^{-1} \left(\begin{array}{c} 0 \\ 1 \\ 0 \end{array}
\right)  $ has only two non-vanishing coefficients,
\BEQ (w_y)_y:=\Big(\Big(\frac{\partial}{\partial\eps_y}
\tilde{\nabla}f\Big) \ (\tilde{\nabla}f)^{-1} \left(\begin{array}{c} 0 \\ 1 \\ 0 \end{array}
\right)  \Big)_y\equiv\frac{1}{{\mathrm{det}}_a^2(u;v)} \gamma_y,
\EEQ
\BEQ \gamma_y:=
\alpha_y \Big( |v|_a^2 \frac{u_y}{a_y}
-(u,v)_a \frac{v_y}{a_y} \Big)+ \beta_y (-2|v|_a^2) 
\EEQ
and
\BEQ (w_y)_{\lambda}:=\Big(\Big(\frac{\partial}{\partial\eps_y}
\tilde{\nabla}f\Big) \ (\tilde{\nabla}f)^{-1} \left(\begin{array}{c} 0 \\ 1 \\ 0 \end{array}
\right)  \Big)_{\lambda}= \frac{1}{{\mathrm{det}}_a^2(u;v)}\beta_y \Big(|v|_a^2 \frac{u_y}{a_y}-
(u,v)_a \frac{v_y}{a_y}\Big).
\EEQ

\item[(ii)] (non-diagonal coefficient) if $x\not=y$, then
\BEA && -\Big( (\tilde{\nabla} f)^{-1}\ \Big(\frac{\partial}{\partial\eps_y}
\tilde{\nabla}f\Big) \ (\tilde{\nabla}f)^{-1} \left(\begin{array}{c} 0 \\ 1 \\ 0 \end{array}
\right)  \Big)_{x} \nonumber\\
&&\qquad=  \frac{1}{{\mathrm{det}}_a^4(u;v)}
\Big[ \half Z_{x,y}\gamma_y- \Big(  |v|_a^2 \frac{u_x}{a_x}
-(u,v)_a \frac{v_x}{a_x} \Big) \beta_y  \Big(  |v|_a^2 \frac{u_y}{a_y}
-(u,v)_a \frac{v_y}{a_y} \Big) \Big]  \nonumber\\
\EEA

\item[(iii)] (diagonal coefficient) letting $x=y$, then
\BEA
 && -\Big( (\tilde{\nabla} f)^{-1}\ \Big(\frac{\partial}{\partial\eps_x}
\tilde{\nabla}f\Big) \ (\tilde{\nabla}f)^{-1} \left(\begin{array}{c} 0 \\ 1 \\ 0 \end{array}
\right)  \Big)_{x} \nonumber\\
&&\qquad=  \frac{1}{{\mathrm{det}}_a^4(u;v)}
\Big[- \half({\mathrm{det}}_a^2(u;v)\,  a_x^{-1}-Z_{x,x})\gamma_x- \Big(  |v|_a^2 \frac{u_x}{a_x}
-(u,v)_a \frac{v_x}{a_x} \Big) \beta_x  \Big(  |v|_a^2 \frac{u_x}{a_x}
-(u,v)_a \frac{v_x}{a_x} \Big) \Big]  \nonumber\\
\EEA

\end{itemize}

We now split accordingly the third line of (\ref{eq:second-derivative}),  
 \BEQ D^2_3:=\sum_x p_x \frac{r_x}{b_x} \ln(\frac{b_x}{r_x})
\ \sum_y 
 \Big((\tilde{\nabla} f)^{-1}  \left(\begin{array}{c} 0 \\ 1 \\ 0 \end{array}\right) \Big)_{y}\ 
 \frac{\partial}{\partial\eps_y}  \Big((\tilde{\nabla} f)^{-1}  \left(\begin{array}{c} 0 \\ 1 \\ 0 \end{array}\right) \Big)_{x}.
  \EEQ
into the sum of  three contributions, 
\BEQ  D_3^2\equiv \frac{1}{{\mathrm{det}}^6_a(u;v)} \big(D_{3,\beta}^2+D_{3,\gamma}^2+D^2_{3,\gamma,diag}
\big).
\EEQ
Note that, now that we are located on the front, we can use (\ref{eq:b*}) and
replace in the sum $ p_x \frac{r_x}{b_x}\ln(b_x/r_x)$ by $mu_x-\frac{\mu}{2}(v_x-2u_x)$ (as in  {\em A.2}), while $\alpha_y,\beta_y$ are
given by (\ref{eq:alphay}), (\ref{eq:betay}).

\Medskip By tedious computations,
we find :
\BEA &&D^2_{3,\beta}=-\sum_{x,y} (mu_x-\frac{\mu}{2}(v_x-2u_x)) 
\Big(  |v|_a^2 \frac{u_x}{a_x}
-(u,v)_a \frac{v_x}{a_x} \Big) \beta_y  \Big(  |v|_a^2 \frac{u_y}{a_y}
-(u,v)_a \frac{v_y}{a_y} \Big)^2 \nonumber\\
&&=\Big\{\sum_y \frac{u^2_y}{p_y}  \Big(  |v|_a^2 \frac{u_y}{a_y}
-(u,v)_a \frac{v_y}{a_y} \Big)^2 \Big\}  \ \Big\{
(m+\mu)\, {\mathrm{det}}_a^2(u;v) \Big\} \nonumber\\
&& =(m+\mu){\mathrm{det}}^2_a(u;v) 
\ \times  \frac{4}{\mu^2} \Big\{ (1+2\mu)\,  (u,v)_a^2 -2(2+\mu)\, (u,v)_a+4\Big\}-\frac{8}{\mu} I_{3,1} \nonumber\\
  \label{eq:D23beta}
\EEA

Similar computations (which we choose to skip) yield $D^2_{3,\gamma}+D^2_{3,\gamma,diag}=0$.  All together, we have found:

\BEQ 
1+\half \frac{d^2(\sigma_W^2)}{dm^2}= \frac{1}{{\mathrm{det}}^2_a(u;v)} \frac{2}{\mu} (2-(u,v)_a)
  \label{eq:th2}
\EEQ



\Bigskip {\em  B. 5. (final formula).}
Finally, from this last formula, we obtain 
\BEA &&  \sigma_W\frac{d^2 \sigma_W}{dm^2} = \half 
\frac{d^2(\sigma_W^2)}{dm^2}-\frac{1}{\sigma_W^2}(\half \frac{d\sigma_W^2}{dm})^2= 
(\ref{eq:th2})-1 - \frac{\mu^2}{\sigma_W^2}  \label{eq:sigmad2sigma}\\
&&= \frac{4/\mu}{{\mathrm{det}}_a^2(u;v)} \Big\{ (
1-\half(u,v)_a) - \frac{\mu}{4} 
{\mathrm{det}}^2_a(u;v) (1+\frac{\mu^2}{\sigma_W^2})\Big\}
\nonumber\\
&&\equiv   \frac{4\mu}{ \sigma_W^2  {\mathrm{det}}_a^2(u;v)}
\, \times\, {\cal D}^2,  \label{eq:calD2}
\EEA
where
\BEQ  {\cal D}^2 :=\frac{{\mathrm{det}}_a^2(u;v)}{4} \sigma_W^3 
\frac{d^2\sigma_W}{dm^2}=
\frac{\sigma_W^2}{\mu^2} \Big[ (
1-\half(u,v)_a)-\frac{\mu}{4} 
{\mathrm{det}}^2_a(u;v) \Big] - \frac{\mu}{4}  {\mathrm{det}}_a^2(u;v)
    \label{eq:sigmad2sigma/dm2} 
\EEQ

Ths implies, finally, Theorem \ref{th:second-derivative}.

\Bigskip{\bf V. 2. 2. Proof of Theorem \ref{th:convexity}.}

\Bigskip  {\em 1. General notations.} For the proof, we first introduce {\em new notations:} 
\BEQ s_x=\frac{p_x}{b_x}, \label{eq:s} \EEQ
 so that
$u_x=r_x s_x$, and $\frac{u_x}{v_x}=\frac{u_x}{2r_x}=\frac{s_x}{2}$; 
recall $\frac{1}{a_x}=\frac{p_x}{u_x(u_x+\frac{\mu}{2}v_x)}$, a formula which
connects scalar products $(\cdot,\cdot)_a$ to standard averages
$\langle \, \cdot\, \rangle$. Also, $\langle \frac{1}{s}\rangle=
\sum_x b_x=1$.  Finally, let 
\BEQ  L_{\mu}:= \frac{1}{\mu}\big(1-\langle \frac{1}{s+\mu}\rangle\big), \qquad S:=\langle \frac{1}{s^2}\rangle=\frac{d (\mu L_{\mu})}{d\mu}\big|_{\mu=0}, \qquad C:=\langle 
\frac{1}{s^3}\rangle=-\half \frac{d^2 (\mu L_{\mu})}{d\mu^2}\big|_{\mu=0}
\EEQ
(S=square, C=cube). Then 
\BEA && \frac{\sigma_W^2}{\mu^2}=\frac{1}{\mu^2} \Big(\langle W^2\rangle -m^2) = \frac{1}{\mu^2} \Big\{  \langle\, \big( m -\mu 
(\frac{b_x}{p_x}-1) \big)^2 \,\rangle - m^2 \Big\} \nonumber\\
&&=  \langle\, (\frac{b_x}{p_x}-1)^2 \, \rangle= 
\langle \frac{1}{s^2}\rangle-1=S-1
\EEA
\BEQ 1-\half(u,v)_a= 1- \langle \frac{1}{s+\mu}\rangle=\mu L_{\mu},
\EEQ
\BEQ 
 1-\half(1+\mu)(u,v)_a=1-(1+\mu) \langle \frac{1}{s+\mu}\rangle=\frac{\mu}{4}{\mathrm{det}}^2_a(u;v) =\mu\big[(1+\mu)L_{\mu}-1\big],
\EEQ
hence $\frac{d^2 \sigma_W}{dm^2}$ has same sign as 
\BEA && \frac{1}{4}  {\mathrm{det}}^2_a(u;v) \frac{{\cal D}^2}{\mu}= \Big\{ (1+\mu)L_{\mu}-1\Big\} \ \Big\{(S-1)L_{\mu}-S((1+\mu)L_{\mu}-1)
\Big\} \nonumber\\
&&\qquad =  \Big\{ (1+\mu)L_{\mu}-1\Big\} \ \Big\{ S-(1+S\mu)L_{\mu}
\Big\} \label{eq:D2}
\EEA

\Bigskip {\em 2. Main Lemma.} The proof of Theorem \ref{th:convexity} rests on the following

\begin{Lemma}  \label{lem:4.3}
\begin{enumerate}
\item
Assume that $\mu>\max\Big(-1, \max\{-s_x, x\in X\}\Big)$ or\\  $\mu<\min\Big(-1,\min\{-s_x,x\in X\} \Big)$.    Then  $(1+\mu)L_{\mu}-1>0$ . 
\item Assume that  $\mu>\max\Big(-1/S,\max\{-s_x, x\in X\}\Big)$ or \\
$\mu<\min\Big(-1/S,\min\{-s_x,x\in X\} \Big)$.
Then $\langle \frac{1+S\mu}{s+\mu}\rangle> 1$.
\end{enumerate}

\end{Lemma}

The conditions of the Lemma hold true in particular (i) if $\mu\ge 0$;
(ii) if $-1\ll \mu<0$ (including in particular the case of the part of the lower front with $\langle w\rangle <0$, as we shall see);
(iii) if $\mu\ll -1$ (including in particular a neighborhood of Kelly's strategy -- where $\mu=-\infty$ --  on the upper front). 

\medskip
{\bf Proof.} 
\begin{enumerate}
\item 
We first note that $\mu((1+\mu)L_{\mu}-1)=(1+\mu) (1-\langle 
\frac{1}{s+\mu}\rangle)-\mu=1-\langle \frac{\mu+1}{\mu+s}\rangle.$
Thus we need to prove that $x(\mu):=\langle \frac{\mu+1}{\mu+s}\rangle< 1$, resp. $> 1$,
for $\mu> 0$, resp. $\mu< 0$. Note first that $x(0)=\langle \frac{1}{s}\rangle=1$.

\Medskip
Let $y(\mu):= \langle (1-s) \frac{\mu+1}{\mu+s}\rangle$.  {\em Mind that
$y$ is considered here as a function of $\mu$ for $b=(b_x)$ fixed. }
 Then
\BEQ \frac{dy}{d\mu} = -
\langle \frac{(1-s)^2}{(\mu+s)^2} \rangle < 0 \label{eq:dy/dmu}
\EEQ
 Since $y(0)=\langle \frac{1}{s}\rangle-1=0$, this means that $y(\mu)$ has  opposite sign
w.r. to $\mu$. When $\mu> 0$, we get
$y(\mu)< 0$, or equivalently,
\BEQ x(\mu)< (1+\mu) \, \langle \frac{s}{s+\mu}\rangle = 
(1+\mu) \big(1- \langle \frac{\mu}{s+\mu}\rangle\big)=  1+\mu-\mu x(\mu) \label{eq:x(mu)<1+mu}
\EEQ
whence $x(\mu)< 1$. When $\max\Big(-1, \max\{-s_x, x\in X\}\Big)<\mu< 0$, we get $y(\mu)> 0$, 
from which $x(\mu)> 1$ .

\Medskip {\em Consider now the case $\mu<\min\Big(-1,\min\{-s_x,x\in X\} \Big)$.} Let  $\mu\to -\infty$, then  $y(-\infty)=1-\langle s\rangle =
1-\langle (\frac{1}{s})^{-1}\rangle \le 1- \langle \frac{1}{s}\rangle^{-1}=0$ by Jensen's inequality (which is coherent with
the previous analysis in the neighborhood of Kelly's strategy, since $y(-\infty)=y(+\infty)<0$). Then  (\ref{eq:dy/dmu}) still holds,
so $y(\mu)<0$, or equivalently, $x(\mu)<1+\mu-\mu x(\mu)$ (see
(\ref{eq:x(mu)<1+mu})), and then (since $\mu<-1$) $x(\mu)>1$.

\item  We let this time $x(\mu):=\langle \frac{S\mu+1}{\mu+s}\rangle$ and $y(\mu):=\langle (S-\frac{1}{s})\frac{S\mu+1}{\mu+s}\rangle$.   {\em As in the previous point, only $\mu$  is varied.}
 Then
\BEQ \frac{dy}{d\mu}=\langle \frac{(Ss-1)(S-\frac{1}{s})}{(\mu+s)^2}
\rangle \ >0   \label{eq:dy/dmubis} 
\EEQ
and $y(0)=0$, hence $y(\mu)$ has same sign as $\mu$. When 
$\mu> 0$, we get $y(\mu)>0$, or equivalently,
\BEA && Sx(\mu) >   \langle \frac{S\mu+1}{s(\mu+s)} \rangle  \nonumber\\
&&=  \langle \frac{1}{s} \frac{S(\mu+s)+1-Ss}{\mu+s} \rangle 
= S+ \frac{1}{\mu} \langle \frac{1}{s}-\frac{1}{\mu+s}\rangle 
- \langle \frac{S}{\mu+s} \rangle \nonumber\\
&&= (S+\frac{1}{\mu}) (1-\langle \frac{1}{\mu+s}\rangle)= S+\frac{1}{\mu} - \frac{x(\mu)}{\mu}  \label{eq:Sx(mu)>S+1/mu}
\EEA
whence $x(\mu)> 1$. When $\max\Big(-1/S,\max\{-s_x, x\in X\}\Big)<\mu<0$, on the other hand, we get 
$y(\mu)<0$, from which $(S+\frac{1}{\mu})x(\mu)< S+\frac{1}{\mu}$, 
whence $(1+\mu S)x(\mu)>1+\mu S$, giving still $x(\mu)>1$.

\Medskip {\em Assume now that $\mu<\min\Big(-1/S,\min\{-s_x,x\in X\} \Big)$.} Let 
$\mu\to -\infty$, then $y(-\infty)=S^2-S>0$ (which is, again, coherent with Kelly's strategy value $y(+\infty)=y(-\infty)>0$).   The inequality
(\ref{eq:dy/dmubis})  holds, whence $y(\mu)>0$, from which
$(S+\frac{1}{\mu})x(\mu)>S+\frac{1}{\mu}$ (see (\ref{eq:Sx(mu)>S+1/mu})), and (since $S+\frac{1}{\mu}>0$), $x(\mu)>1$ still.
\end{enumerate}

\Bigskip{\em Proof of Theorem \ref{th:convexity}.} 
The second derivative of the curve has same sign (see (\ref{eq:D2})) as 
\BEQ F(\mu):=\Big\{ (1+\mu)L_{\mu}-1\Big\} \ \Big\{ S-(1+S\mu)L_{\mu}
\Big\} .
 \EEQ Now,
\BEA &&  F(\mu)= -\frac{(1+\mu)L_{\mu}-1}{\mu} \Big\{(1+S\mu )(1-\langle 
\frac{1}{s+\mu}\rangle)- S\mu  \Big\}
\nonumber\\
&&\qquad =  \Big\{ \frac{(1+\mu)L_{\mu}-1}{\mu} \Big\}\ \Big\{ \langle \frac{1+S\mu}{s+\mu}\rangle  -1 \Big\}
\EEA
We now use Lemma \ref{lem:4.3} (2) and {\em assume first that $\mu>0$}. When $\mu>0$, $\langle \frac{1+S\mu}{s+\mu}\rangle>1$ so $F(\mu)>0$; thus that part of the 
front  is strictly convex. Now (see eq.
\ref{eq:58} and eq. (\ref{eq:sigmad2sigma})),  $\half \frac{d\mu}{dm}=\half \frac{d^2(\sigma_W^2)}{dm^2} \ge 
\sigma_W \frac{d^2\sigma_W}{dm^2}>0$, whence $\frac{d\langle W\rangle}{d\mu}>0$.
Starting from $\mu=0^+$ and increasing $\mu$, one thus moves through
the part of the lower front where $\langle W\rangle>0$.

\Medskip  {\em Assume now that  $\mu<\min\Big(-1,\min\{-s_x,x\in X\} \Big)$}. Then   $F(\mu)<0$; that  part of the  front is strictly concave. This holds true in particular in a neighborhood of 
Kelly's point on the upper front. 

\Medskip {\em Assume finally that  $0>\mu>\max\Big(-1/S,\max\{-s_x, x\in X\}\Big)$.} Then $F(\mu)<0$.  That part of the front is strictly concave.
This holds true in particular (as seen using the same argument
as in the case $\mu>0$)  in a neighborhood of the null strategy on the left, i.e. when $\langle W \rangle<0$.

\section{Horse race simulation}
For a given set of odds $o_x$, probabilities $p_x$ and gambler's betting strategy $b_x$,  
simulations represented in Fig. 1 in the main text have been performed by following these steps:%
\begin{enumerate}
  \item Simulation starts with unit capital $C_0=1$
  \item A random number $r\in (0,1)$ is generated with a uniform distribution. 
  \item If $r<p_1$, horse number $1$ wins this race. Otherwise, if $p_1<r<p_1+p_2$ then horse 2 wins this race. Otherwise, if $p_1+p_2<r<p_1+p_2+p_3$ horse 3 wins the race and so on. This assures that 
  every horse wins a race according to the $p_x$ distribution.
  \item Capital of the gambler is updated, according to the bets which have been placed and 
  according to the odds given by bookmaker (i.e. using equation 1 in the main text).
  \item Steps 2-4 are repeated for the number of desired races.
  \item The log of the capital evolution is displayed in the figure.
\end{enumerate}
The different lines in Figure 1 correspond to different realizations of the previous steps, 
with different betting strategies $b_x$ but the same odds and probabilities. We have simulated a race with 
three horses with $o_1=4$ ($r_1=1/4$), $o_2=4$ ($r_2=1/4$), and $p_1=0.1$, $p_2=0.3$.

\section{Numerical optimization}

To find the optimal bets $b^*$ for every objective function and parameter values, we use a {\em simulated annealing} algorithm, useful for global optimization problems in large search spaces, specially to avoid local maxima solutions. Starting from an initial guess $b_0$, a neighboring valid bets vector is randomly generated. It is accepted with probability 1 if it produces an increase in the value of the objective function $J$, that is, an ``uphill'' move. If not, then it is only accepted with probability proportional to an exponential factor $\exp(-\beta J)$, where $\beta$ plays the role of an inverse temperature. The algorithm is then repeated, increasing the value of $\beta$ in each iteration. In the beginning, with a small value for $\beta$ ``uphill'' moves are allowed, although with less probability than ``downhill'' moves, which avoids getting trapped in a local maximum. After a number of iterations, a stable value for the bets vector is obtained that gives an approximation to the optimal one.

As explained in the main text, there is a phase transition in this model where 
the optimal strategy changes from the null strategy to a mixed strategy when 
$\gamma$ approaches $\gamma_c$. Because of this transition, 
the Pareto front near the null strategy has a triangular shape, with a slope determined by 
this critical $\gamma_c$ as shown in Fig. \ref{figS1}.
\begin{figure}[h]
\includegraphics[scale=0.45]{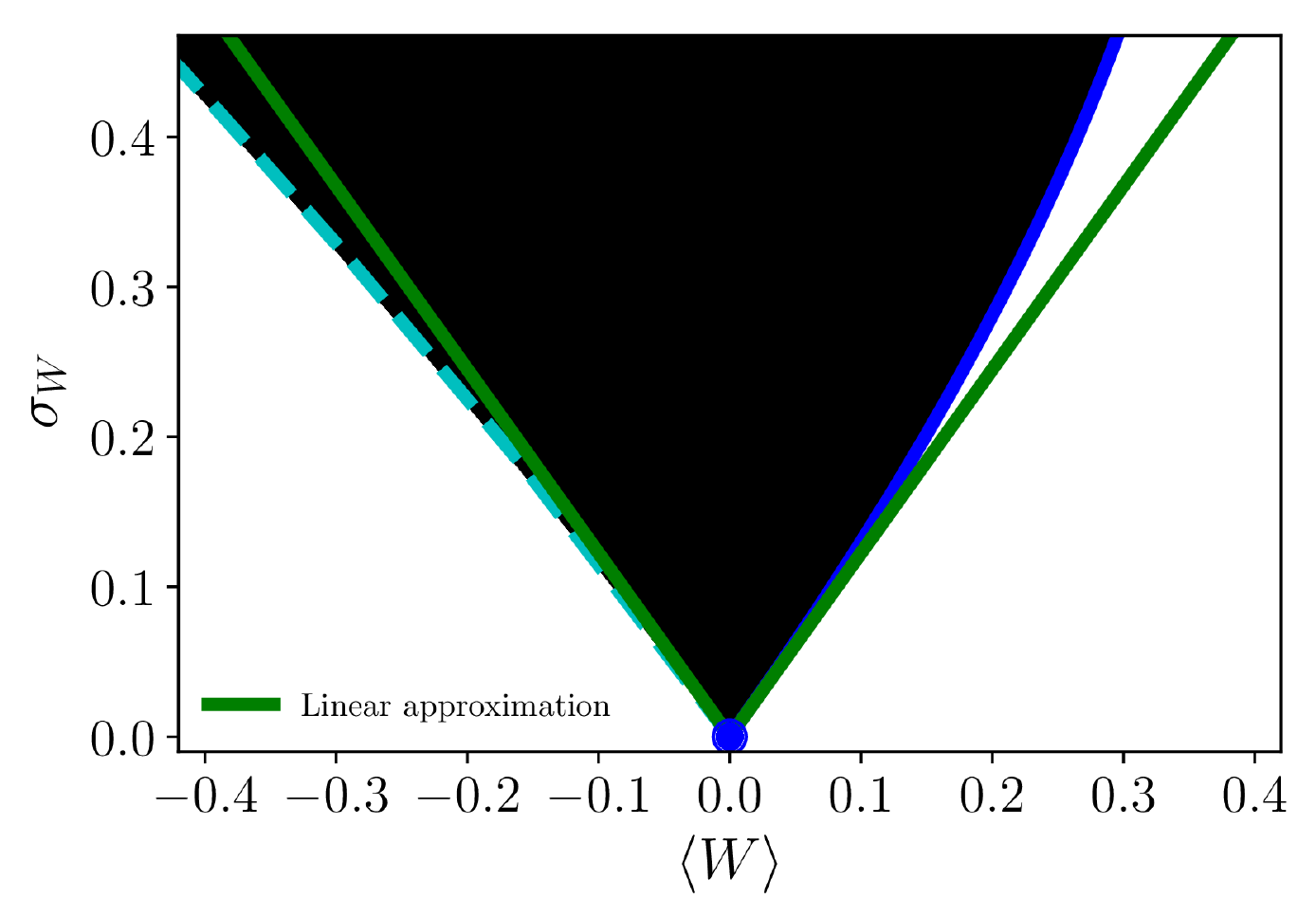} 
\caption{Zoom of the Pareto front shown in Fig 3 of the main text near the null strategy. The green solid lines  correspond to the linear approximation used near the null strategy and the black points forming a dense region inside the front have been randomly chosen just as in Fig 3. The figure corresponds to the case of three horses in the uncorrelated case with parameters $p_1=0.2,p_2=0.4,r_1=0.4,r_2=0.2$.} 
\label{figS1}
\end{figure}

The last figure of the main text, namely Fig 4, represents the Pareto border for 3 horses in the correlated case. 
We provide here some details on the parameters used to make this figure.
Let $P(i,j)=P(i|j)$ represents the conditional probabilities of horse $i$ winning provided horse $j$ won in the previous round. Then $P$ is 3 by 3 matrix, which we choose to be
\begin{equation*}
P = 
\begin{pmatrix}
0.2 & 0.2 & 0.3 \\
0.4 & 0.5 & 0.4 \\
0.4 & 0.3 & 0.3
\end{pmatrix}.
\end{equation*}
The odds of the first two horses are
$o_1=4$ ($r_1=1/4$), $o_2=4$ ($r_2=1/4$),
while the parameters of the last horse are deduced by normalization of the bets and of the probabilities.


\end{widetext}
 

\end{document}